\RequirePackage{fix-cm}
\documentclass[twocolumn]{svjour3}          
\smartqed  
\usepackage{mathptmx}      
\usepackage{graphicx}
\usepackage{amsmath,amsfonts,amssymb}
\usepackage{bm} 
\PassOptionsToPackage{hyphens}{url}\usepackage[pdftex]{hyperref}
\usepackage{ulem} 
\usepackage{color}
\usepackage{multirow,array}
\usepackage[linesnumbered,ruled,vlined]{algorithm2e}
\SetKwInput{KwInput}{Input}
\SetKwInput{KwOutput}{Output}
\allowdisplaybreaks

\journalname{Gran. Mat.}

\begin{document}

\title{Contact model for elastically anisotropic bodies and efficient implementation into the discrete element method}

\titlerunning{Contact model for elastically anisotropic bodies}        

\author{Saviz Mowlavi \and Ken Kamrin}

\institute{Saviz Mowlavi \at
              Department of Mechanical Engineering, Massachusetts Institute of Technology, Cambridge, MA, USA \\
              \email{smowlavi@mit.edu}           
              \and
              Ken Kamrin \at
              Department of Mechanical Engineering, Massachusetts Institute of Technology, Cambridge, MA, USA \\
              \email{kkamrin@mit.edu}  
}

\date{Received: date / Accepted: date}

\maketitle

\begin{abstract}
We introduce a contact law for the normal force generated between two contacting, elastically anisotropic bodies of arbitrary geometry. The only requirement is that their surfaces be smooth and frictionless. This anisotropic contact law is obtained from a simplification of the exact solution to the continuum elasticity problem and takes the familiar form of Hertz' contact law, with the only difference being the orientation-dependence of the material-specific contact modulus. The contact law is remarkably accurate when compared with the exact solution, for a wide range of materials and surface geometries. We describe a computationally efficient implementation of the contact law into a discrete element method code, taking advantage of the precomputation of the contact modulus over all possible orientations. Finally, we showcase two application examples based on real materials where elastic anisotropy of the particles induces noticeable effects on macroscopic behavior.
\end{abstract}

\section{Introduction}

Beginning with the seminal paper of Cundall and Strack \cite{cundall1979}, the Discrete Element Method (DEM) has rapidly established itself as a method of choice for simulating the behavior of granular materials in a wide range of situations \cite{zhu2008,luding2008,guo2015}. In this approach, Newton's equations of motion are integrated individually for every particle in the system, taking into account body forces as well as surface forces that arise from the interactions of contacting particles. Contact force laws dictate the magnitude of these surface forces as a function of the overlap between adjacent particles. As such, they are an essential ingredient of any DEM simulation, and a multitude of contact laws of various complexities have been formulated to account for effects as varied as friction \cite{cundall1979,walton1986,thornton1991}, damping \cite{brilliantov1996}, torsion \cite{dintwa2005}, cohesion \cite{soulie2006,lim2006,marshall2009}, plasticity \cite{tomas2007}, and so forth.

Contact force models may be divided into two broad classes \cite{zhu2007}. The first concerns models that are formulated based on an exact or approximate solution of the physics governing the contact problem at the scale of the individual grains. The most prominent example is Hertz' contact law \cite{hertz1882}, which gives an expression for the normal force generated by the elastic deformation of two contacting spheres. Hertz' contact law, which is based on the exact solution of the continuum elasticity equations for this problem, takes a remarkably simple form wherein the force is dependent on the three halves power of the overlap distance between the particles \cite{johnson1987}. Contact force models belonging to the second class are formulated empirically, balancing ease of implementation and computational cost with accuracy of the results. Cundall and Strack's linear spring-dashpot model falls under this second category.

Although the second class of methods is particularly useful when one wants to incorporate physical mechanisms that elude simple analytical solutions, the first class is preferable when one is concerned with the precise quantification of the forces in a granular medium. For instance, numerous studies \cite{mueth1998,howell1999,clark2012} have investigated the distribution and properties of interparticle forces in granular materials and their connection with the external loading characteristics.
In the case of elastically isotropic bodies, for which the contact force is independent of the direction of contact, Hertz' contact law will return the exact forces as long as the deformation of the bodies is small and contact points on the same particle are not too close. Most materials in nature, however, are elastically anisotropic. Whenever the size of individual particles becomes small enough, crystalline grains become apparent relative to the particle size \cite{du2017}, and the contact force between particles will be direction-dependent as a result of elastic anisotropy. Myriad engineering processes such as additive manufacturing \cite{meier2019} or ceramic packings \cite{hang2017,crystal2020} involve powders of fine particles, and the accurate quantification of interparticle forces in these cases calls for a contact force law that can take elastic anisotropy into account. Clearly, one expects van der Waals forces to play an important role at these small scales, but the relative strength of such attractive forces decreases with increasing load and their modeling has already been treated previously \cite{johnson1971,barber2014}. Furthermore, monocrystalline granular particles of a larger size do exist, both naturally \cite{klein2002} and artificially \cite{dhanaraj2010}. Such particles notably play a key role in novel experimental methods for inferring particle-wise strain tensors in opaque packings by exploiting X-ray diffraction \cite{hall2015,hurley2016}.

In this paper, we derive a contact law for the normal elastic force that is generated between two elastically anisotropic bodies of arbitrary geometry, as long as the surfaces are smooth and frictionless. Our approach begins with the formulation of a numerical procedure for the exact analytical solution to the continuum elasticity problem, which builds on more than fifty years of research in the contact mechanics literature \cite{barber2000}. In particular, several authors have sought to extract a relationship between indentation force, depth, and contact area during the unloading branch of an indentation test, wherein an axisymmetric rigid indentor is pressed against an elastically anisotropic half space \cite{vlassak1993,vlassak1994,swadener2001,vlassak2003,delafargue2004}. The exact solution procedure that we present here extends the scope of these studies to the case of two contacting, elastically anisotropic bodies with smooth and non-spherical geometry, which lacks a detailed treatment in the previous endeavors.

We then simplify the exact solution into a readily implementable anisotropic contact force law, which in the particular case of spherical contacting bodies $B_1$ and $B_2$ of radii $R^{B_1}$ and $R^{B_2}$ takes the form
\begin{equation}
F = \frac{4}{3} \tilde{E}_*^c(\alpha^{B_1},\beta^{B_1},\alpha^{B_2},\beta^{B_2}) R^{1/2} \delta^{3/2},
\label{eq:AnisotropicLawIntro}
\end{equation}%
where $F$ is the normal force and $\delta$ the overlap between the two bodies, $R = (1/R^{B_1} + 1/R^{B_2})^{-1}$ is the composite radius, $\tilde{E}_*^c$ is a material-specific composite modulus depending on two sets of Euler angles $(\alpha^B,\beta^B)$ describing the orientation of the contact normal direction with respect to the internal axes of bodies $B = B_1$ and $B_2$. The \textit{only} difference between the simplified anisotropic contact law \eqref{eq:AnisotropicLawIntro} and Hertz' familiar contact law for isotropic bodies lies in the orientation-dependence of the composite modulus $\tilde{E}_*^c$, which calls upon the entire set of elastic constants for the material comprising each body. This similarity between the isotropic and simplified anisotropic contact laws extends to smooth particles of abritrary shape, as we show later in the paper.

The simplification utilizes Vlassak \textit{et al.}'s \cite{vlassak2003} idea of truncating the Fourier series expansion of the surface Green's function to its constant term, which was shown in \cite{vlassak2003} to result in accurate predictions of the force generated by a rigid spherical indentor on a half space made of sapphire. We demonstrate that this accuracy is retained in our contact law for generic smooth contacting particles over a wide range of materials, with the simplified anisotropic contact law differing from the exact solution by less than 1\% in all considered cases. We then show how to efficiently implement these formulas in a DEM scheme, taking advantage of the offline precomputation of the material-specific contact modulus over all possible orientations. Finally, we present two examples involving assemblies of single-crystal zirconia particles that display how anisotropy at the particle level alters macroscopic behavior and can be exploited in applications.

The paper is organized as follows. We begin by formulating the contact problem and describe the solution methodology in Section \ref{sec:ProblemStatement}, based on which an exact anisotropic contact force law is then derived in Section \ref{sec:HertzianSolution}. Through successive simplifications of the exact solution, we then propose in Section \ref{sec:PossibleSimplifications} two simplified contact force laws, which we compare with the exact solution in Section \ref{sec:ComparisonContactForceLaws} after validating the latter against finite-element simulations. An implementation of the first simplified contact law into a DEM code is then presented in Section \ref{sec:Applications}, along with two example applications featuring elastically anisotropic particles. Conclusions close the paper in Section \ref{sec:Conclusions}. Finally, we invite the reader interested in the implementation details to consult the appendices.

\section{Problem statement and solution methodology}
\label{sec:ProblemStatement}

\subsection{Definition of the contact problem}

\begin{figure*}
\centering
\includegraphics[width=\textwidth]{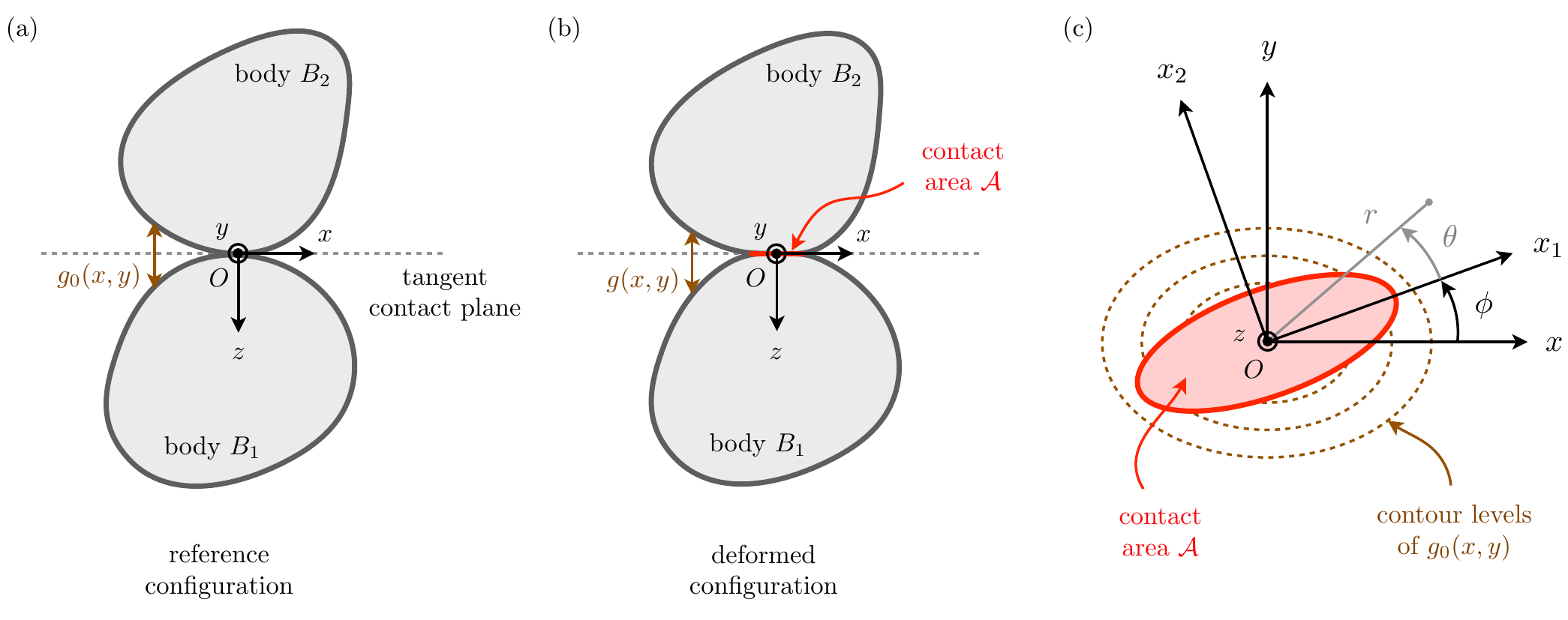}
\caption{Geometry of the contact problem. (a) In the reference configuration, the two bodies are contacting at a single point $O$ and separated by a common tangent plane. The coordinate system $(x,y,z)$ is defined in such a way that the $(x,y)$-axes, spanning the tangent plane, are aligned with the principal axes of the contour levels of the initial gap function $g_0(x,y)$. (b) In the deformed configuration, the two bodies are pressed against each other with a normal force $F$, resulting in a relative displacement normal to the tangent plane as well as the establishment of a finite contact region $\mathcal{A}$. (c) In the tangent contact plane, the contact area takes the shape of an ellipse whose major and minor axes $(x_1,x_2)$ are rotated by an angle $\phi$ with respect to the $(x,y)$ axes. The set of polar coordinates $(r,\theta)$ used in Section \ref{sec:AnisotropicBodies} is defined with respect to the $(x_1,x_2)$ axes.}
\label{fig:Geometry}
\end{figure*}

We consider two elastically anisotropic bodies $B_1$ and $B_2$, comprised of materials having elasticity tensors $\mathbb{C}^{B_1}$ and $\mathbb{C}^{B_2}$. Throughout the paper, quantities with a superscript $B_1$ and $B_2$ will refer to body $B_1$ and body $B_2$, respectively, and quantities with a superscript $B$ will refer to either body interchangeably. In the reference unstressed configuration, the two bodies are contacting at a single point and are separated by a common tangent contact plane, as pictured in Figure \ref{fig:Geometry}(a). 
Let the contact point $O$ be the origin of a cartesian coordinate system $(x,y,z)$, where the $x$-$y$ plane is the common tangent plane and the $z$-axis is directed along the inward normal of body $B_1$. The initial gap $g_0(x,y)$ measures the gap between the undeformed bodies, and is given to lowest order by
\begin{equation}
g_0(x,y) = Mx^2 + Ny^2,
\label{eq:GapFunction}
\end{equation}
where the $x$- and $y$-axes have been chosen so that they align with the principal axes of the contour levels of $g_0(x,y)$, and $N \ge M$ by convention. In this work, we only consider bodies with a smooth and convex surface, for which the first-order terms of $g_0(x,y)$ are zero and $M$, $N$ are both positive. While outside the scope of this paper, we mention that formulae to obtain $M$ and $N$ from the principal radii of curvature of bodies $B_1$ and $B_2$ at the contact point are given in the books of Johnson \cite{johnson1987} and Barber \cite{barber2018}. In the specific case of ellipsoidal bodies, the calculation of the principal radii of curvature knowing the contact point and orientations of $B_1$ and $B_2$ is nontrivial and explained in the appendix of \cite{zheng2013}.

The bodies are then pressed against each other with a force $F$ directed along the normal to the contact plane, which results in a relative displacement $\delta$ and the establishment of a finite contact region $\mathcal{A}$, as pictured in Figure \ref{fig:Geometry}(b). We denote the vertical surface displacement generated in each body along the $z$-axis by $w^{B_1}(x,y)$ and $w^{B_2}(x,y)$, both measured positive into the respective body. The final gap $g(x,y)$ is then given by
\begin{equation}
g(x,y) = g_0(x,y) - \delta + w^{B_1}(x,y) + w^{B_2}(x,y).
\end{equation}
Inside the contact region $\mathcal{A}$, the gap $g(x,y)$ must vanish, which implies
\begin{equation}
w^{B_1}(x,y) + w^{B_2}(x,y) = \delta - g_0(x,y), \quad (x,y) \in \mathcal{A}.
\label{eq:BC1}
\end{equation}
Outside the contact region, the gap $g(x,y)$ must be positive, which translates as
\begin{equation}
w^{B_1}(x,y) + w^{B_2}(x,y) > \delta - g_0(x,y), \quad (x,y) \notin \mathcal{A}.
\label{eq:BC2}
\end{equation}
We assume that the surfaces are frictionless, so that there is only a  normal traction (that is, a pressure) $p(x,y)$ between the bodies, which resultant over the contact area $\mathcal{A}$ is equal to $F$. The boundary conditions \eqref{eq:BC1} and \eqref{eq:BC2} are supplemented by the condition that $p(x,y) > 0$ for $(x,y) \in \mathcal{A}$, and $p(x,y) = 0$ for $(x,y) \notin \mathcal{A}$. 

The problem, therefore, is to find the contact area $\mathcal{A}$ and pressure distribution $p(x,y)$ such that the resulting surface displacements satisfy the boundary conditions \eqref{eq:BC1} and \eqref{eq:BC2}. In this way, the normal force $F$ between the two bodies can be related with their relative displacement $\delta$. This elasticity problem was first solved analytically by Hertz \cite{hertz1882} for elastically isotropic bodies, leading to the well-known Hertz contact law. The solution process is, however, much more cumbersome for elastically anisotropic bodies. While integral expressions have been derived and solution strategies have been suggested by various authors using a range of mathematical techniques \cite{willis1966,barber1992,swadener2001,vlassak2003,gao2007,barber2014}, an exact step-by-step solution scheme for generally-shaped contacting surfaces, including the non-circular case $M \neq N$, is still missing.


\subsection{Solution methodology}

Our general solution strategy for the contact problem is based on Hertz's derivation of the elastically isotropic case \cite{hertz1882,johnson1987}, and proceeds in a similar way for both isotropic and anisotropic bodies. First, one introduces the simplification that the contact region $\mathcal{A}$ is flat, and that the surface displacements generated by the pressure distribution $p(x,y)$ are equal to those that would be produced in equivalent semi-infinite bodies (i.e., elastic half-spaces) loaded with the same surface pressure distribution over the same contact region. In order for this simplification to hold, the size of the contact area must be small with respect to the dimensions of each body as well as their principal radii of curvature at the contact point. This simplification, first introduced by Hertz, enables one to express the combined surface displacements as the convolution 
\begin{multline}
w^{B_1}(x,y) + w^{B_2}(x,y) \\ 
= \sum_{B \in \{B_1,B_2\}} \iint_\mathcal{A} \hat{w}^B(x-x',y-y') p(x',y') dx' dy',
\label{eq:SurfaceDisplacementConvolution}
\end{multline}
where $\hat{w}^B(x-x',y-y')$ is the vertical surface displacement at $(x,y)$ produced by a unit concentrated normal load at $(x',y')$ on the surface of an elastic half-space. As we will see later, the Green's function $\hat{w}^B(x,y)$ is a known quantity that depends on the elasticity tensor $\mathbb{C}^B$ of body $B$ together with, for anisotropic bodies, its orientation with respect to the contact plane.

The next step is to find the shape of the contact region and distribution of pressure such that the combined surface displacements predicted by \eqref{eq:SurfaceDisplacementConvolution} agree with the boundary conditions \eqref{eq:BC1} and \eqref{eq:BC2}. Consider a flat elliptical\footnote{Hertz was guided by his observations of elliptic optical interference fringes between two contacting glass lenses, which is the very problem that motivated his subsequent analysis of the contact deformation \cite{johnson1987}.} contact area with semi-axes lengths $a_1$ and $a_2$,
\begin{equation}
\mathcal{A} = \left\{ (x_1,x_2) : \frac{x_1^2}{a_1^2} + \frac{x_2^2}{a_2^2} < 1 \right\},
\label{eq:ContactArea}
\end{equation}
where $a_2 \le a_1$ by convention, and the $(x_1,x_2)$ coordinates are rotated by some yet-unknown angle $\phi$ about the $(x,y)$ coordinates, as shown in Figure \ref{fig:Geometry}(c). In addition, 
consider a pressure distribution of the form
\begin{equation}
p(x_1,x_2) = p_0 \left(1 - \frac{x_1^2}{a_1^2} - \frac{x_2^2}{a_2^2} \right)^\xi, \quad (x_1,x_2) \in \mathcal{A},
\label{eq:PressureDistribution}
\end{equation}
where the exponent $\xi$ is unknown\footnote{Asymptotic arguments, however, require that for smooth contacting bodies the contact pressure tend to zero at the boundary of the contact area \cite{barber2018}, which implies that $\xi $ is positive. } in advance. For the respective cases of isotropic and anisotropic bodies, Hertz \cite{hertz1882} and Willis \cite{willis1966} showed that when $\xi = 1/2$ (and only then), the postulated contact area \eqref{eq:ContactArea} and pressure distribution \eqref{eq:PressureDistribution} produce combined surface displacements \eqref{eq:SurfaceDisplacementConvolution} that are compatible with the conditions \eqref{eq:BC1} and \eqref{eq:BC2}, thereby validating the functional forms \eqref{eq:ContactArea} and \eqref{eq:PressureDistribution}. In fact, in the isotropic case, it can be immediately shown that \eqref{eq:ContactArea} and \eqref{eq:PressureDistribution} solve (\ref{eq:BC1}-\ref{eq:SurfaceDisplacementConvolution}) by appealing to a known analogous result from potential theory (see \cite{landau1970} for details).

The problem is now reduced to finding the scalar parameters $a_1$, $a_2$, $\phi$, and $p_0$, given $M$, $N$, $\delta$, as well as the orientation and elastic constants of the contacting bodies. Once this is done by equating the coefficients in \eqref{eq:BC1} and \eqref{eq:SurfaceDisplacementConvolution}, the contact law for the force $F$ can be obtained through the relation%
\begin{equation}
F = \iint_\mathcal{A} p(x_1',x_2') dx_1' dx_2' = \frac{2}{3} \pi p_0 a_1 a_2.
\label{eq:ResultantPressure}
\end{equation}
The following section goes through this process in detail, beginning with elastically isotropic bodies in Section \ref{sec:IsotropicBodies} and followed by anisotropic bodies in Section \ref{sec:AnisotropicBodies}. In each case, the Green's function is first presented (equations \eqref{eq:GreenFunctionIsotropic} and \eqref{eq:GreenFunctionAnisotropic} for isotropic and anisotropic bodies, respectively) and inserted in the convolution integral \eqref{eq:SurfaceDisplacementConvolution} to obtain the surface displacements (equations \eqref{eq:SurfaceDisplacementIsotropic} and \eqref{eq:SurfaceDisplacementAnisotropic} for isotropic and anisotropic bodies, respectively). The unknown scalar parameters are then found by enforcing the boundary condition \eqref{eq:BC1}, eventually leading to a relation between the contact force $F$ and the relative displacement $\delta$ (equations \eqref{eq:IsotropicForce} and \eqref{eq:FinalAnisotropicSolution} for isotropic and anisotropic bodies, respectively).

\section{Derivation of the exact contact force}
\label{sec:HertzianSolution}

\subsection{Isotropic bodies}
\label{sec:IsotropicBodies}

\subsubsection{Green's function and surface displacements}
\label{sec:IsotropicGreenSurfaceDisplacements}

We begin with a review of the solution for elastically isotropic bodies, which we will later refer to when developing a simplified anisotropic solution. In the isotropic case, the Green's function $\hat{w}^B(x_1,x_2)$ is axisymmetric and given in closed form as
\begin{equation}
\hat{w}^B(x_1,x_2) = \frac{1}{\pi E_*^B (x_1^2+x_2^2)^{1/2}},
\label{eq:GreenFunctionIsotropic}
\end{equation}
where $E_*^B$ is the plane strain modulus of body $B$, defined from its Young's modulus $E^B$ and Poisson's ratio $\nu^B$ as 
\begin{equation}
E_*^B = \frac{E^B}{1-(\nu^B)^2}.
\end{equation}
Inserting \eqref{eq:GreenFunctionIsotropic} into \eqref{eq:SurfaceDisplacementConvolution} and using \eqref{eq:ResultantPressure}, we find that the combined surface displacement within the contact area $\mathcal{A}$ caused by the pressure distribution \eqref{eq:PressureDistribution} is \cite{barber2018}
\begin{multline}
w^{B_1}(x_1,x_2) + w^{B_2}(x_1,x_2) \\
= \frac{3 F}{4 \pi a_1 E_*^c} \left( I_0(e) - \frac{x_1^2}{a_1^2} I_1(e) - \frac{x_2^2}{a_1^2} I_2(e) \right),
\label{eq:SurfaceDisplacementIsotropic}
\end{multline}
where $E_*^c$ is the composite plane strain modulus, 
\begin{equation}
E_*^c = \left(\frac{1}{E_*^{B_1}} + \frac{1}{E_*^{B_2}}\right)^{-1},
\label{eq:CompositePlaneStrainModulus}
\end{equation} 
$e$ is the eccentricity of the contact area,
\begin{equation}
e = \sqrt{1-\left(\frac{a_2}{a_1}\right)^2},
\end{equation}
and $I_0(e)$, $I_1(e)$, and $I_2(e)$ are integrals defined as
\begin{subequations}
\begin{gather}
I_0(e) = \int_0^{\pi} \frac{d\theta}{(1-e^2 \cos^2 \theta)^{1/2}}, \\
I_1(e) = \int_0^{\pi} \frac{\sin^2 \theta d\theta}{(1-e^2 \cos^2 \theta)^{3/2}}, \\
I_2(e) = \int_0^{\pi} \frac{\cos^2 \theta d\theta}{(1-e^2 \cos^2 \theta)^{3/2}}.
\end{gather}
\label{eq:IntegralsIsotropic}%
\end{subequations}
It now remains to identify the surface displacements \eqref{eq:SurfaceDisplacementIsotropic} with the boundary condition \eqref{eq:BC1} in order to solve for the unknowns $a_1$, $e$, $\phi$, and $F$. This last step of the solution process is described hereafter.

\subsubsection{Contact force solution}
\label{sec:SolutionProcedureIsotropic}

The solution procedure presented here is similar to that given in Barber \cite{barber2018}, with the exception that the latter reference uses complete elliptic integrals of the first and second kind instead of \eqref{eq:IntegralsIsotropic}. This leads to a numerically ill-posed problem when $e$ vanishes, which we avoid by working with expressions \eqref{eq:IntegralsIsotropic}. 

Equating the surface displacements \eqref{eq:SurfaceDisplacementIsotropic} with the boundary condition \eqref{eq:BC1}, we find that the pressure distribution \eqref{eq:PressureDistribution} gives the correct surface displacements provided that the $(x_1,x_2)$ axes coincide with $(x,y)$ (that is, $\phi = 0$, which means that the major and minor axes of the pressure distribution are aligned with those of the initial gap function). In addition, this yields the relations
\begin{subequations}
\begin{gather}
\frac{3 F I_0(e)}{4 \pi a_1 E_*^c} = \delta, \label{eq:Equation1Isotropic} \\
\frac{3 F I_1(e)}{4 \pi a_1^3 E_*^c} = M, \label{eq:Equation2Isotropic} \\
\frac{3 F I_2(e)}{4 \pi a_1^3 E_*^c} = N. \label{eq:Equation3Isotropic}
\end{gather} \label{eq:RelationsIsotropic}%
\end{subequations}
By combining \eqref{eq:Equation2Isotropic} and \eqref{eq:Equation3Isotropic}, we obtain a simple nonlinear equation for the eccentricity,
\begin{equation}
\frac{I_2(e)}{I_1(e)} - \frac{N}{M} = 0.
\label{eq:IsotropicEccentricity}
\end{equation}
The contact force $F$ then follows from \eqref{eq:Equation1Isotropic} and \eqref{eq:Equation2Isotropic} as
\begin{equation}
F = \frac{4 \pi}{3} E_*^c \frac{[I_1(e)]^{1/2}}{[I_0(e)]^{3/2}} M^{-1/2} \delta^{3/2},
\label{eq:IsotropicForce}
\end{equation}
where the material parameter $E_*^c$ is defined in \eqref{eq:CompositePlaneStrainModulus}. In summary, the isotropic contact force law requires the solution of equation \eqref{eq:IsotropicEccentricity} for $e$, after which $F$ can be obtained with \eqref{eq:IsotropicForce}.

\subsubsection{Spherical case}
\label{sec:SolutionProcedureIsotropicSpherical}

We conclude our review of isotropic materials with a discussion on the form of the Hertzian solution for the limiting case of spherical contacting bodies, which results in the celebrated expression commonly referred to as the Hertz contact law \cite{zhu2007,kruggel2007}. Consider two contacting spheres of radii $R^{B_1}$ and $R^{B_2}$. To lowest order, the gap between the undeformed bodies is given by
\begin{equation}
g_0(x,y) = \frac{x^2}{2R} + \frac{y^2}{2R},
\end{equation}
where $1/R = 1/R^{B_1} + 1/R^{B_2}$. Therefore $M = N = 1/2R$, in which case \eqref{eq:IsotropicEccentricity} gives $e = 0$, and \eqref{eq:IsotropicForce} reduces to the Hertz contact law,
\begin{equation}
F = \frac{4}{3} E_*^c R^{1/2} \delta^{3/2}.
\label{eq:HertzContactLaw}
\end{equation}

\subsection{Anisotropic bodies}
\label{sec:AnisotropicBodies}

\subsubsection{Green's function and surface displacements}
\label{sec:AnisotropicGreenSurfaceDisplacements}

\begin{figure*}
\centering
\includegraphics[width=\textwidth]{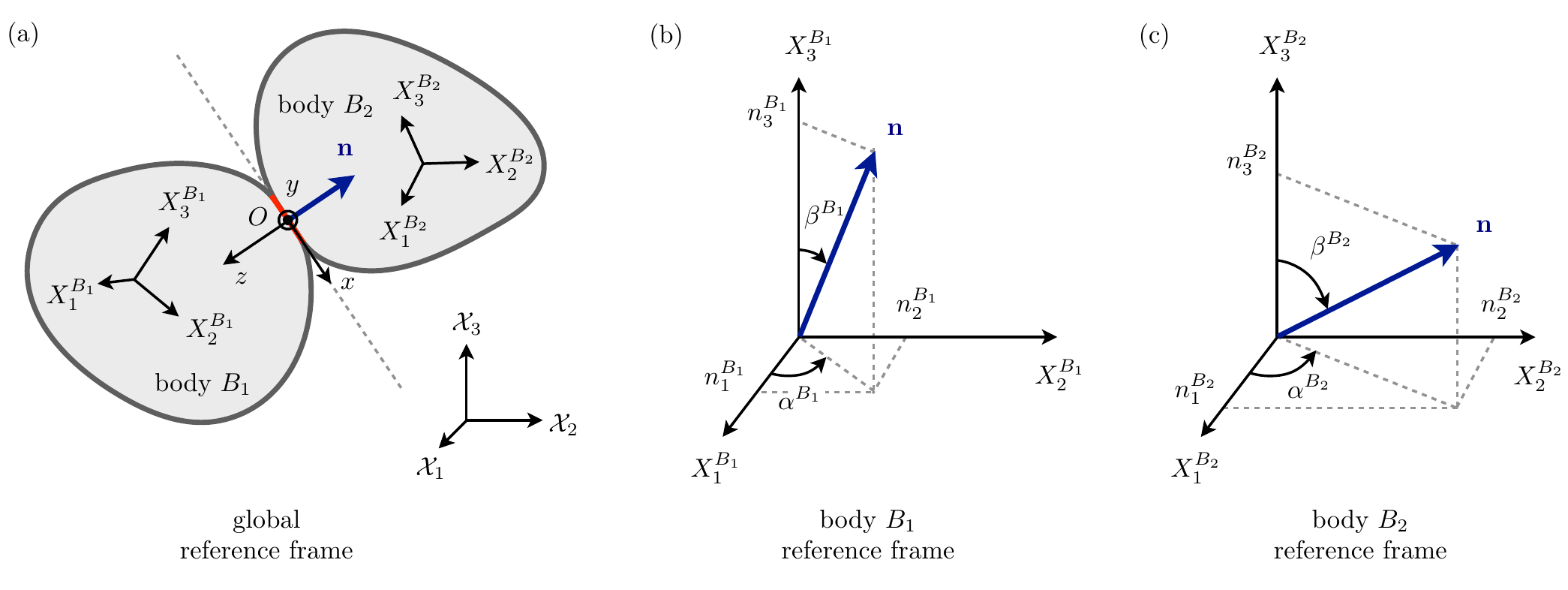}
\caption{Further details on the geometry of the problem. (a) The two contacting bodies depicted in Figure \ref{fig:Geometry} are here viewed from the global reference frame, defined by the coordinates $(\mathcal{X}_1,\mathcal{X}_2,\mathcal{X}_3)$. We represent the local coordinates bases $(X_1^{B_1},X_2^{B_1},X_3^{B_1})$ and $(X_1^{B_2},X_2^{B_2},X_3^{B_2})$ of bodies $B_1$ and $B_2$, the contact normal and tangent plane directions $(x,y,z)$, as well as the unit normal $\mathbf{n}$ to the tangent contacting plane. (b,c) In the reference frames $(X_1^{B_1},X_2^{B_1},X_3^{B_1})$ and $(X_1^{B_2},X_2^{B_2},X_3^{B_2})$ of body $B_1$ and $B_2$, respectively, the unit-length contact normal $\mathbf{n}$ can be parameterized either by its coordinates $(n_1^B,n_2^B,n_3^B)$, or by the two Euler angles $(\alpha^B,\beta^B)$.}
\label{fig:GeometryBody}
\end{figure*}%

For anisotropic bodies, there is no direct algebraic expression for the Green's function $\hat{w}^B(x_1,x_2)$. Various integral expressions have been derived by different authors, starting with Willis \cite{willis1966} who performed a Fourier transform in the $x$-$y$ plane and solved implicitly the resulting equations. Willis' expression, however, requires the simultaneous solution of multiple nonlinear integral equations, making it challenging to work with in practice. Instead, we utilize in this paper a direct integral expression for the Green's function derived by Barnett and Lothe \cite{barnett1975}, obtained by solving the Fourier-transformed equations using a formalism due to Stroh \cite{stroh1958}.

First, let the coordinates $(X_1^B,X_2^B,X_3^B)$ represent a basis that is preferentially oriented for the material structure in body $B$, and with respect to which the components of the elasticity tensor are $\mathbb{C}_{ijkm}^B$. In body $B$, the stress and strain are therefore everywhere related as
\begin{equation}
\epsilon_{ij}^B = \mathbb{C}_{ijkm}^B \sigma_{km}^B,
\end{equation}
where $\sigma_{km}^B$ and $\epsilon_{ij}^B$ are, respectively, the components of the local stress and strain tensors in the $(X_1^B,X_2^B,X_3^B)$ basis. Figure \ref{fig:GeometryBody}(a) depicts this body-centric basis for the same bodies $B_1$ and $B_2$ introduced in Figure \ref{fig:Geometry}, but here viewed from the global reference frame $(\mathcal{X}_1,\mathcal{X}_2,\mathcal{X}_3)$.
(The latter is introduced for future reference and will not be referred to in this section.) We introduce the unit normal $\mathbf{n}$ to the contact plane, which is directed from body $B_1$ to body $B_2$, i.e.~along the negative $z$-direction. As pictured in Figures \ref{fig:GeometryBody}(b) and \ref{fig:GeometryBody}(c), we denote by $(n_1^B,n_2^B,n_3^B)$ the components of $\mathbf{n}$ in the $(X_1^B,X_2^B,X_3^B)$ basis of each body. Then, Barnett and Lothe's expression for the vertical displacement at a point $P$ in the $x$-$y$ plane due to a concentrated unit vertical load at the origin reads\footnote{Although the unit normal $\mathbf{n}$ is shared between bodies $B_1$ and $B_2$ and hence points in opposite directions with respect to each body's surface, expression \eqref{eq:GreenFunctionAnisotropic} is valid for both bodies since it is quadratic in the components of $\mathbf{n}$.} (see the appendix of \cite{vlassak1994})
\begin{equation}
\hat{w}^B(\mathbf{x}) = \frac{1}{|\mathbf{x}|} \left[ n_k^B G_{km}^{-1}\left(\frac{\mathbf{x}}{|\mathbf{x}|}\right) n_m^B \right],
\label{eq:GreenFunctionAnisotropic}
\end{equation}
where $\mathbf{x}$ is the position vector of $P$. The matrix $[\mathbf{G}]$ in the above equation is defined as
\begin{equation}
G_{ij}(\mathbf{t}) = \int_0^{2\pi} \left( \{\mathbf{r} \mathbf{r}\}_{ij} - \{\mathbf{r} \mathbf{s}\}_{ik} \{\mathbf{s} \mathbf{s}\}_{kr}^{-1} \{\mathbf{s} \mathbf{r}\}_{rj} \right) d\gamma,
\label{eq:GreenIntegral}
\end{equation}
where $\mathbf{r}, \mathbf{s}, \mathbf{t}$ are unit vectors such that $(\mathbf{r}, \mathbf{s}, \mathbf{t})$ forms a right-hand Cartesian system, $\gamma$ is the angle between $\mathbf{r}$ and some fixed point in the plane perpendicular to $\mathbf{t}$, and the matrices $(\mathbf{a} \mathbf{b})$ are defined as
\begin{equation}
\{\mathbf{a} \mathbf{b}\}_{jk} = a_i \mathbb{C}_{ijkm}^B b_m,
\label{eq:GreenIntegralMatrices}
\end{equation}
with $(a_1,a_2,a_3)$ and $(b_1,b_2,b_3)$ denoting the components of vectors $\mathbf{a}$ and $\mathbf{b}$ in the $(X_1^B,X_2^B,X_3^B)$ basis. It now remains to substitute \eqref{eq:GreenFunctionAnisotropic} into \eqref{eq:SurfaceDisplacementConvolution} and solve the resulting integral. This is no easy task, but Barber and Ciavarella \cite{barber2014} have suggested an efficient strategy for doing so, which we formalize here. 

We define the set of polar coordinates $(r,\theta)$ as $(x_1,x_2) = (r \cos \theta, r \sin \theta)$, as shown in Figure \ref{fig:Geometry}(c). The angle $\theta$ is measured with respect to the $(x_1,x_2)$ axes, which are rotated by an as-yet-unknown angle $\phi$ with respect to the $(x,y)$ axes. Since the orientation of the latter with respect to the $(X_1^B,X_2^B,X_3^B)$ basis is known, we write the Green's function \eqref{eq:GreenFunctionAnisotropic} in the `rotated' polar coordinates $(r,\theta)$ as
\begin{equation}
\hat{w}^B(r,\theta;\phi) = \frac{1}{r} \left[ n_k^B G_{km}^{-1}(\theta;\phi) n_m^B \right] = \frac{h^B(\theta;\phi)}{r},
\label{eq:GreenFunctionAnisotropicPolar}
\end{equation}
where the presence of $\phi$ emphasizes the dependence of the polar Green's function on the orientation $\phi$ of the $(x_1,x_2)$ basis. For completeness, we shall mention that $h^B(\theta;\phi)$ is also a function of the material parameters as well as the (known) orientation of the $(x,y,z)$ basis with respect to the $(X_1^B,X_2^B,X_3^B)$ basis, which can be characterized by a rotation matrix as described in Appendix \ref{app:CoordinateTransformations}. In an effort to preserve clarity of exposure, however, we have omitted this dependence in our notation. 

As a consequence of Maxwell's reciprocal theorem (see \cite{barber2018}), the function $h^B(\theta;\phi)$ satisfies the relation $h^B(\theta;\phi) = h^B(\theta+\pi;\phi)$ and therefore admits the Fourier expansion
\begin{equation}
h^B(\theta;\phi) = \sum_{m = 0}^\infty a_m^B(\phi) \cos 2m\theta + \sum_{m = 1}^\infty b_m^B(\phi) \sin 2m\theta.
\label{eq:GreenFunctionAnisotropicFourier}
\end{equation}
Due to the way that the angles $\theta$ and $\phi$ are defined in Figure \ref{fig:Geometry}(c), we necessarily have $h^B(\theta;\phi) = h^B(\theta+\phi;0)$. As a consequence, the Fourier coefficients $a_m^B(\phi)$ and $b_m^B(\phi)$ can be expressed as
\begin{subequations}
\begin{align}
a_m^B(\phi) &= a_m^B(0) \cos 2m\phi + b_m^B(0) \sin 2m\phi, \\
b_m^B(\phi) &= -a_m^B(0) \sin 2m\phi + b_m^B(0) \cos 2m\phi.
\end{align}
\label{eq:RotationFourierCoefficients}%
\end{subequations}
Therefore, the knowledge of $h^B(\theta;0)$ suffices to calculate the Fourier coefficients $a_m^B(\phi)$ and $b_m^B(\phi)$. Given the elasticity tensor $\mathbb{C}^B$ as well as the orientation of the $(x,y,z)$ basis with respect to the $(X_1^B,X_2^B,X_3^B)$ basis, we present in Appendix \ref{app:CalculationGreenFunction} an algorithm for computing $h^B(\theta;0)$. In practice, the Fourier coefficients $a_m^B(0)$ and $b_m^B(0)$ decay very quickly with $m$, and we have found that truncating the Fourier series at $m = 5$ is perfectly adequate. 

As shown in Barber and Ciavarella \cite{barber2014}, the integral \eqref{eq:SurfaceDisplacementConvolution} can then be solved in polar coordinates using \eqref{eq:GreenFunctionAnisotropicFourier}, leading to the combined surface displacement
\begin{align}
w^{B_1}(x_1,&x_2) + w^{B_2}(x_1,x_2) = \nonumber \\ 
&\frac{3F}{4a_1} \bigg\{ \sum_{m=0}^\infty a_m(\phi) \left[ I_{0,m}(e) - \frac{x_1^2}{a_1^2} I_{1,m}(e) - \frac{x_2^2}{a_1^2} I_{2,m}(e) \right] \nonumber \\ 
&+ \frac{x_1 x_2}{a_1^2} \sum_{m=1}^\infty b_m(\phi) I_{3,m}(e) \bigg\},
\label{eq:SurfaceDisplacementAnisotropic}
\end{align}
where $a_m(\phi) = a_m^{B_1}(\phi) + a_m^{B_2}(\phi)$, $b_m(\phi) = b_m^{B_1}(\phi) + b_m^{B_2}(\phi)$, and the integrals $I_{0,m}(e)$, $I_{1,m}(e)$, $I_{2,m}(e)$, and $I_{3,m}(e)$ are defined as
\begin{subequations}
\begin{gather}
I_{0,m}(e) = \int_0^{\pi} \frac{\cos(2m\theta) d\theta}{(1-e^2 \cos^2 \theta)^{1/2}}, \\
I_{1,m}(e) = \int_0^{\pi} \frac{\sin^2 \theta \cos(2m\theta) d\theta}{(1-e^2 \cos^2 \theta)^{3/2}}, \\
I_{2,m}(e) = \int_0^{\pi} \frac{\cos^2 \theta \cos(2m\theta) d\theta}{(1-e^2 \cos^2 \theta)^{3/2}}, \\
I_{3,m}(e) = \int_0^{\pi} \frac{\sin(2\theta) \sin(2m\theta) d\theta}{(1-e^2 \cos^2 \theta)^{3/2}}.
\end{gather}
\end{subequations}
Note that these integrals relate to the ones defined in \eqref{eq:IntegralsIsotropic} for isotropic bodies as $I_{0,0}(e) = I_0(e), I_{1,0}(e) = I_1(e)$, and $I_{2,0}(e) = I_2(e)$. Finally, identifying the surface displacement \eqref{eq:SurfaceDisplacementAnisotropic} with the boundary condition \eqref{eq:BC1}, one can solve for $a_1$, $e$, $\phi$, and $F$. This requires an iterative approach which we describe hereafter.

\subsubsection{Contact force solution}

We now present a solution procedure that goes beyond the solutions detailed in \cite{vlassak2003} and \cite{barber2014}, which are restricted to the specific case $M = N$. First, we express the initial gap function $g_0$ in the $(x_1,x_2)$ coordinates; see Figure \ref{fig:Geometry}(c). This can be done by substituting the coordinate transformation relations
\begin{subequations}
\begin{align}
x = x_1 \cos \phi - x_2 \sin \phi, \\
y = x_2 \cos \phi + x_1 \sin \phi,
\end{align}
\end{subequations}
into \eqref{eq:GapFunction}, leading to
\begin{multline}
g_0 = x_1^2(M \cos^2 \phi + N \sin^2 \phi) \\ 
+ x_2^2(M \sin^2 \phi + N \cos^2 \phi) + x_1 x_2 (N-M) \sin 2\phi.
\end{multline}
Equating the surface displacements \eqref{eq:SurfaceDisplacementAnisotropic} with the boundary condition \eqref{eq:BC1} in the $(x_1,x_2)$ coordinates, we obtain the relations
\begin{subequations}
\begin{align}
\frac{3F}{4a_1} \sum_{m=0}^\infty a_m(\phi) I_{0,m}(e) &= \delta, \label{eq:Equation1} \\
\frac{3F}{4a_1^3} \sum_{m=0}^\infty a_m(\phi) I_{1,m}(e) &= M \cos^2 \phi + N \sin^2 \phi, \label{eq:Equation2} \\
\frac{3F}{4a_1^3} \sum_{m=0}^\infty a_m(\phi) I_{2,m}(e) &= M \sin^2 \phi + N \cos^2 \phi, \label{eq:Equation3} \\
\frac{3F}{4a_1^3} \sum_{m=1}^\infty b_m(\phi) I_{3,m}(e) &= (M-N) \sin 2\phi. \label{eq:Equation4}
\end{align}%
\end{subequations}
We recast \eqref{eq:Equation2} to \eqref{eq:Equation4} into two equations for $\phi$ and $e$:
\begin{subequations}
\begin{gather}
(M-N) \sin 2\phi \sum_{m=0}^\infty a_m(\phi) I_{2,m}(e) \nonumber \\ 
\hspace{20pt} - (M \sin^2 \phi + N \cos^2 \phi) \sum_{m=1}^\infty b_m(\phi) I_{3,m}(e) = 0, \label{eq:OffsetAngleEccentricityRelation1} \\
(M \cos^2 \phi + N \sin^2 \phi) \sum_{m=0}^\infty a_m(\phi) I_{2,m}(e) \nonumber \\
\hspace{20pt} - (M \sin^2 \phi + N \cos^2 \phi) \sum_{m=0}^\infty a_m(\phi) I_{1,m}(e) = 0. \label{eq:OffsetAngleEccentricityRelation2}
\end{gather}
\label{eq:OffsetAngleEccentricityRelations}%
\end{subequations}
Together, \eqref{eq:OffsetAngleEccentricityRelation1} and \eqref{eq:OffsetAngleEccentricityRelation2} form a nonlinear system of equations for $e$ and $\phi$ that can be solved numerically according to the procedure described in Appendix \ref{app:EccentricityPhaseSolution}, after which the only remaining unknowns are $a_1$ and $F$. Combining \eqref{eq:Equation1} and \eqref{eq:Equation2}, we find that $F$ is given by
\begin{equation}
F = \frac{4}{3} \frac{[\sum_{m=0}^\infty a_m(\phi) I_{1,m}(e)]^{1/2}}{[\sum_{m=0}^\infty a_m(\phi) I_{0,m}(e)]^{3/2}} (M \cos^2 \phi + N \sin^2 \phi)^{-1/2} \delta^{3/2}.
\label{eq:FinalAnisotropicSolution}
\end{equation}
In summary, the anisotropic contact force law requires the calculation of $h^B(\theta;0)$ from equation \eqref{eq:GreenFunctionAnisotropicPolar}, after which the Fourier coefficients $a_m^B(\phi)$ and $b_m^B(\phi)$ can be found using \eqref{eq:GreenFunctionAnisotropicFourier} and \eqref{eq:RotationFourierCoefficients}. These can then be substituted into equations \eqref{eq:OffsetAngleEccentricityRelation1} and \eqref{eq:OffsetAngleEccentricityRelation2} to calculate $e$ and and $\phi$, before finally obtaining $F$ through equation \eqref{eq:FinalAnisotropicSolution}. Observe that the anisotropic solution retains the power $3/2$ dependence of $F$ on $\delta$ from the isotropic solution \eqref{eq:IsotropicForce}. Moreover, in the limiting case of isotropic materials, one obtains $a_0^B(\phi) = 1/\pi E_*^B$ and $a_m^B(\phi) = b_m^B(\phi) = 0$ for all $m > 0$, and this solution procedure appropriately reduces to the isotropic one given in Section \eqref{sec:SolutionProcedureIsotropic}.

\subsubsection{Limitations for an implementation in DEM}

We end this section with a discussion on issues of computational cost. While the solution procedure presented in this section is reasonably fast so long as one is merely interested in computing the force between two bodies under a few different situations, it is nevertheless too expensive for direct implementation into a DEM code. Indeed, the latter case requires a calculation of the force at every contact and at every time step, in which case the solution scheme quickly becomes prohibitively expensive.  An alternative option is to precompute, for a given material, a look-up table of stored solution values for $e$ and $\phi$ that would then be accessed during the course of the DEM simulation, with only the force $F$ remaining to compute from \eqref{eq:FinalAnisotropicSolution}. However, such a table would have to be four-dimensional -- three parameters to describe the orientation of the $(x,y,z)$ basis with respect to the $(X_1^B,X_2^B,X_3^B)$ basis, and one for the ratio $N/M$ -- due to the coupling between equations \eqref{eq:OffsetAngleEccentricityRelations} for $e$ and $\phi$, and the Fourier coefficients of the Green's function. In practice, this is not possible from a storage requirement standpoint, which essentially precludes the applicability of the exact contact force law \eqref{eq:FinalAnisotropicSolution} to the DEM. In order to circumvent this issue, we discuss in the following section two possible simplification strategies, which both rely on shortening the form of the Green's function \eqref{eq:GreenFunctionAnisotropicPolar} appearing in the exact solution. We also propose an efficient implementation of the simplified solutions into DEM simulations.

\section{Simplifications of the anisotropic contact force}
\label{sec:PossibleSimplifications}

\subsection{Isotropic truncation of the Green's function}
\label{sec:Approximation1}

This approximation follows the exact anisotropic solution detailed in Section \ref{sec:AnisotropicBodies}, with the crucial difference that the Fourier expansion \eqref{eq:GreenFunctionAnisotropicFourier} of the Green's function \eqref{eq:GreenFunctionAnisotropicPolar} is truncated after the constant term $a_0^B(\phi)$, so that $a_m^B(\phi) = b_m^B(\phi) = 0$ for all $m > 0$. This idea of truncating the Green's function was introduced by Vlassak \textit{et al.} \cite{vlassak2003} in the context of a rigid indentor pressing against an anisotropic half space.

Setting $m = 0$ in \eqref{eq:RotationFourierCoefficients} reveals that $a_0^B$ is not a function of $\phi$, as expected since the constant term is equal to the average of $h^B(\theta;\phi)$ over all $\theta$. In contrast to the other Fourier coefficients, it follows that $a_0^B$ no longer depends on the full orientation of the $(x,y,z)$ basis with respect to the $(X_1^B,X_2^B,X_3^B)$ basis attached to body $B$, but only on the orientation of the unit contact normal $\mathbf{n}$ with respect to $(X_1^B,X_2^B,X_3^B)$. As shown in Figures \ref{fig:GeometryBody}(b) and \ref{fig:GeometryBody}(c), this relative orientation can be parameterized either by the components $(n_1^B,n_2^B,n_3^B)$ of $\mathbf{n}$ or by the two Euler angles $(\alpha^B,\beta^B)$, both measured with respect to the local $(X_1^B,X_2^B,X_3^B)$ basis. The two representations are related as
\begin{subequations}
\begin{align}
(n_1^B,n_2^B,n_3^B) &= (\cos \alpha^B \sqrt{1-\cos^2 \beta^B}, \nonumber \\ 
&\quad \ \ \sin \alpha^B \sqrt{1-\cos^2 \beta^B}, \cos \beta^B), \label{eq:FromEulerToComponents} \\
(\alpha^B, \beta^B) &= (\text{arctan2}(n_2^B,n_1^B),\arccos n_3^B), \label{eq:FromComponentsToEuler}
\end{align}
\end{subequations}
where $\text{arctan2}(\cdot,\cdot)$ denotes the four-quadrant inverse tangent. From here on, we will indicate the contact normal direction with respect to body $B$ in terms of the Euler angles $(\alpha^B,\beta^B)$, and the dependence of $a_0^B$ on the latter will be denoted explicitly.

After truncation of the Fourier series, the Green's function \eqref{eq:GreenFunctionAnisotropicPolar} reduces to the same form as that for isotropic bodies,
\begin{equation}
\hat{w}^B(r) = \frac{a_0^B(\alpha^B,\beta^B)}{r} = \frac{1}{\pi \tilde{E}_*^B(\alpha^B,\beta^B) r},
\label{eq:GreenFunctionIsotropicTruncation}
\end{equation}
where $\tilde{E}_*^B(\alpha^B,\beta^B)$ is the plane strain modulus of the equivalent isotropic body, defined by Vlassak \textit{et al.}~\cite{vlassak2003} as
\begin{equation}
\tilde{E}_*^B(\alpha^B,\beta^B) = \frac{1}{\pi a_0^B(\alpha^B,\beta^B)}.
\label{eq:EquivalentPlaneStrainModulus}
\end{equation}
In \eqref{eq:GreenFunctionIsotropicTruncation} and \eqref{eq:EquivalentPlaneStrainModulus}, the superscript $B$ attached to $a_0^B$ and $\tilde{E}_*^B$ indicates a dependence of these quantities on the elasticity tensor $\mathbb{C}^B$ of body $B$, which may differ between bodies $B_1$ and $B_2$.

By virtue of the similarity between the truncated Green's function \eqref{eq:GreenFunctionIsotropicTruncation} and its isotropic counterpart \eqref{eq:GreenFunctionIsotropic}, the rest of our solution proceeds in an analogous way to isotropic materials and is considerably simpler than the full anisotropic solution. Similar to the isotropic case detailed in Section \ref{sec:SolutionProcedureIsotropic}, the phase angle $\phi = 0$ and the eccentricity $e$ satisfies
\begin{equation}
\frac{I_2(e)}{I_1(e)} - \frac{N}{M} = 0.
\label{eq:IsotropicEccentricityApprox}
\end{equation}
Hence, the normal force $F$ is expressed as
\begin{equation}
F = \frac{4 \pi}{3} \tilde{E}_*^c(\alpha^{B_1},\beta^{B_1},\alpha^{B_2},\beta^{B_2}) \frac{[I_1(e)]^{1/2}}{[I_0(e)]^{3/2}} M^{-1/2} \delta^{3/2},
\label{eq:IsotropicForceApprox}
\end{equation}
where $\tilde{E}_*^c$, the composite plain strain modulus of the equivalent isotropic bodies, is given by
\begin{multline}
\tilde{E}_*^c(\alpha^{B_1},\beta^{B_1},\alpha^{B_2},\beta^{B_2}) \\
= \left(\frac{1}{\tilde{E}_*^{B_1}(\alpha^{B_1},\beta^{B_1})} + \frac{1}{\tilde{E}_*^{B_2}(\alpha^{B_2},\beta^{B_2})}\right)^{-1}.
\label{eq:EquivalentCompositePlaneStrainModulus}
\end{multline}

\subsubsection{Spherical case}
\label{sec:SolutionProcedureAnisotropicSpherical}

Similarly to isotropic materials, the particular case of spherical bodies lends itself to further simplification. As discussed in Section \ref{sec:SolutionProcedureIsotropicSpherical}, the gap function coefficients for two contacting spheres of radii $R^{B_1}$ and $R^{B_2}$ are given by $M = N = 1/2R$, with $1/R = 1/R^{B_1} + 1/R^{B_2}$. It then follows that the eccentricity $e = 0$, and the normal force $F$ reduces to
\begin{equation}
F = \frac{4}{3} \tilde{E}_*^c(\alpha^{B_1},\beta^{B_1},\alpha^{B_2},\beta^{B_2}) R^{1/2} \delta^{3/2}.
\label{eq:IsotropicForceApproxCircular}
\end{equation}

\subsubsection{Efficient implementation in DEM through a look-up table}
\label{sec:EfficientImplementationInDEM}

The simplified solutions \eqref{eq:IsotropicForceApprox} and \eqref{eq:IsotropicForceApproxCircular} obtained from the truncation of the Green's function assume the same form as the exact isotropic solutions \eqref{eq:IsotropicForce} and \eqref{eq:HertzContactLaw}, with the exception of $\tilde{E}_*^c$, the composite plain strain modulus \eqref{eq:EquivalentCompositePlaneStrainModulus}. In the anisotropic solution, the latter depends on the relative orientation of the contact normal with respect to the two bodies through the equivalent plane strain modulus $\tilde{E}_*^B(\alpha^B,\beta^B)$ defined in \eqref{eq:EquivalentPlaneStrainModulus}. The computation of $\tilde{E}_*^B(\alpha^B,\beta^B)$ through the truncation of the Green's function \eqref{eq:GreenFunctionAnisotropicPolar} is rather demanding, which prevents its online integration into a DEM code. Nevertheless, we may leverage the fact that besides the angles $\alpha^B$ and $\beta^B$, the quantity $\tilde{E}_*^B$ solely depends on the elasticity tensor $\mathbb{C}^B$ of body $B$. 

An effective remedy to the computational cost issue is thus to create, for every different material $\mathbb{C}^B$ present in the simulation, a table $[\tilde{E}_*](\cdot,\cdot\,;\mathbb{C}^B)$ of values of the equivalent plane strain modulus spanning all contact normal directions $\alpha^B \in [0,2\pi]$ and $\beta^B \in [0,\pi]$. These two-dimensional look-up tables are to be precomputed offline and their values interpolated online according to the instantaneous values of $\alpha^B$ and $\beta^B$ when \eqref{eq:EquivalentCompositePlaneStrainModulus} is called during the course of the DEM simulation. In this way, the simplified anisotropic contact laws \eqref{eq:IsotropicForceApprox} and \eqref{eq:IsotropicForceApproxCircular} are equally fast to compute as their isotropic counterparts, save for the interpolation of the look-up tables. Given a material, we describe in Appendix \ref{app:CalculationPSM} an algorithm for the calculation of such a look-up table --- this table is then \textit{shared among all bodies made of the same material}. The value of the composite plain strain modulus $\tilde{E}_*^c$ corresponding to two contacting bodies $B_1$ and $B_2$ can then be retrieved from two (or one, if $\mathbb{C}^{B_1} = \mathbb{C}^{B_2}$) precomputed tables $[\tilde{E}_*](\cdot,\cdot\,;\mathbb{C}^{B_1})$ and $[\tilde{E}_*](\cdot,\cdot\,;\mathbb{C}^{B_2})$ according to the algorithm presented in Appendix \ref{app:CalculationCompositePSM}.

Lastly, the solution to \eqref{eq:IsotropicEccentricityApprox} for the eccentricity $e$ of the contact area, which is required for non-spherical particles in both the isotropic and anisotropic contact laws, takes just a few Newton-Raphson iterations to converge\footnote{To speed up convergence, one may start the iterations from $e = 2e_g/\sqrt{3}$ with $e_g = \sqrt{1-M/N}$, which provides an excellent approximation to the solution in the range $0 < e_g < 0.4$ and remains reasonably accurate up to $e_g \simeq 0.8$ (see \cite{barber2018}, Section 3.3.1).} and can either be directly implemented into a DEM code, or stored in another one-dimensional look-up table as a function of the ratio $M/N$.

With $\tilde{E}_*^c$ and $e$ in hand, the contact force can be readily calculated from \eqref{eq:IsotropicForceApprox}. A Python implementation of the computational approach described in this section, including the computation of the look-up table, has been shared in an online repository at \url{https://github.com/smowlavi/AnisotropicGrains.git}.

\subsection{Ad hoc computation of the plane strain modulus}
\label{sec:Approximation2}

As we have noted above, the main issue with the first simplification strategy lies in the need to compute the anisotropic Green's function \eqref{eq:GreenFunctionAnisotropicPolar} in order to obtain the equivalent plane strain modulus $\tilde{E}_*^B(\alpha^B,\beta^B)$ defined in \eqref{eq:EquivalentPlaneStrainModulus}. In this section, we present an alternative, ad hoc approach to obtain $\tilde{E}_*^B$ that is much faster to compute, yet retains directional information and makes full use of all elastic constants of the material. Recall that for isotropic materials, the plain strain modulus is given by
\begin{equation}
E_*^B = \frac{E^B}{1-(\nu^B)^2},
\end{equation}
where $E^B$ and $\nu^B$ are respectively the Young's modulus and Poisson's ratio of body $B$. Returning to anisotropic materials, we may define an ad hoc equivalent plain strain modulus $\tilde{E}_*^B(\alpha^B,\beta^B)$ through a direct generalization of the above expression. We substitute $E^B$ and $\nu^B$ with the \textit{effective} Young's modulus $E_\mathbf{n}^B(\alpha^B,\beta^B)$ and \textit{effective} Poisson's ratio $\nu_\mathbf{n}^B(\alpha^B,\beta^B)$ along the contact normal direction $\mathbf{n}$, giving
\begin{equation}
\tilde{E}_*^B(\alpha^B,\beta^B) = \frac{E_\mathbf{n}^B(\alpha^B,\beta^B)}{1-(\nu_\mathbf{n}^B(\alpha^B,\beta^B))^2}.
\label{eq:EffectivePlaneStrainModulus}
\end{equation} 
The effective material quantities $E_\mathbf{n}^B$ and $\nu_\mathbf{n}^B$ are defined the same way as for isotropic materials, with the exception that they now depend on the relative orientation $(\alpha^B,\beta^B)$ of the unit normal $\mathbf{n}$ with respect to the body. First, consider a state of uniform uniaxial stress along $\mathbf{n}$,
\begin{equation}
\bm{\sigma} = \sigma \mathbf{n} \otimes \mathbf{n},
\end{equation}
which induces a strain $\bm{\epsilon}^B = \mathbb{S}^B \bm{\sigma}$, with $\mathbb{S}^B$ the compliance tensor of particle $B$. The resulting normal strain along the contact normal $\mathbf{n}$ is then given by
\begin{equation}
\epsilon_\mathbf{n}^B = \mathbf{n} \cdot \bm{\epsilon}^B \mathbf{n} = \mathbf{n} \cdot (\mathbb{S}^B \bm{\sigma}) \mathbf{n},
\end{equation}
and the resulting normal strain in the transverse direction is given by
\begin{equation}
\epsilon_\mathbf{t}^B = \frac{1}{2\pi} \int_0^{2\pi} (\mathbf{t} \cdot \bm{\epsilon}^B \mathbf{t}) d\gamma = \frac{1}{2\pi} \int_0^{2\pi} (\mathbf{t} \cdot (\mathbb{S}^B \bm{\sigma}) \mathbf{t}) d\gamma,
\end{equation}
where $\mathbf{t}$ is a unit vector orthogonal to $\mathbf{n}$, and $\gamma$ is the angle between $\mathbf{t}$ and an arbitrary fixed point in the plane perpendicular to $\mathbf{n}$. Denoting $\mathbf{u}$, $\mathbf{v}$ a fixed orthogonal basis within that plane, the substitution $\mathbf{t} = \cos \gamma \, \mathbf{u} + \sin \gamma \, \mathbf{v}$ enables the explicit calculation of the above integral, leading to
\begin{equation}
\epsilon_\mathbf{t}^B = \frac{1}{2} (\mathbf{u} \cdot (\mathbb{S}^B \bm{\sigma}) \mathbf{u}) + \frac{1}{2} (\mathbf{v} \cdot (\mathbb{S}^B \bm{\sigma}) \mathbf{v}).
\end{equation}
The effective Young's modulus and Poisson's ratio along $\mathbf{n}$ are thus
\begin{equation}
E_\mathbf{n}^B(\alpha^B,\beta^B) = \frac{\sigma}{\epsilon_\mathbf{n}^B}, \qquad \nu_\mathbf{n}^B(\alpha^B,\beta^B) = - \frac{\epsilon_\mathbf{t}^B}{\epsilon_\mathbf{n}^B}.
\end{equation}
Finally, we insert the above quantities back into the ad hoc definition \eqref{eq:EffectivePlaneStrainModulus} of the equivalent plane strain modulus, and we use \eqref{eq:IsotropicEccentricityApprox} and \eqref{eq:IsotropicForceApprox} to find the resulting normal force. We note that while the computation of the plane strain modulus using the ad hoc approach described here is much faster than the Green's function approach described in Section \ref{sec:Approximation1}, it is still more demanding than simply retrieving a precomputed value from a look-up table. Therefore, it is also advantageous to use the latter approach in this case, creating a table of values $[\tilde{E}_*](\cdot,\cdot\,;\mathbb{C}^B)$ of the ad hoc plain strain modulus as a function of $\alpha^B$ and $\beta^B$, for every material present in the simulation.

\subsection{Summary of the exact and simplified laws}

\begin{table*}
\begin{center}
\renewcommand{\arraystretch}{1.2}
\begin{tabular}{|r|cccc|cccc|}
\hline
\multirow{2}{*}{Contact force law} & \multicolumn{4}{c|}{Elliptic $g_0$ ($M \neq N$)} & \multicolumn{4}{c|}{Circular $g_0$ ($M = N$)} \\ \cline{2-9}
					    & $F$ & $\tilde{E}_*^B$ & $e$ & $\phi$ & $F$ & $\tilde{E}_*^B$ & $e$ & $\phi$ \\ \hline
Exact (Section \ref{sec:AnisotropicBodies}) & \eqref{eq:FinalAnisotropicSolution} & -- & \eqref{eq:OffsetAngleEccentricityRelations} & \eqref{eq:OffsetAngleEccentricityRelations} & \eqref{eq:FinalAnisotropicSolution} & -- & \eqref{eq:OffsetAngleEccentricityRelations} & \eqref{eq:OffsetAngleEccentricityRelations} \\
Truncated (Section \ref{sec:Approximation1}) & \eqref{eq:IsotropicForceApprox} & \eqref{eq:EquivalentPlaneStrainModulus}  & \eqref{eq:IsotropicEccentricityApprox} & 0 & \eqref{eq:IsotropicForceApproxCircular} & \eqref{eq:EquivalentPlaneStrainModulus} & 0 & 0 \\
Ad hoc (Section \ref{sec:Approximation2}) & \eqref{eq:IsotropicForceApprox} & \eqref{eq:EffectivePlaneStrainModulus} & \eqref{eq:IsotropicEccentricityApprox} & 0 & \eqref{eq:IsotropicForceApproxCircular} & \eqref{eq:EffectivePlaneStrainModulus} & 0 & 0 \\ \hline
\end{tabular}
\caption{Summary of the exact and simplified anisotropic contact laws.}
\label{tab:SummaryContactLaws}
\end{center}
\end{table*}

For the convenience of the reader, we provide in Table \ref{tab:SummaryContactLaws} a summary of the exact and simplified anisotropic contact force laws that we have presented in Sections \ref{sec:AnisotropicBodies}, \ref{sec:Approximation1}, and \ref{sec:Approximation2}. We display separately the general case of an elliptic gap function (i.e.~$M \neq N$) and the limiting case of a circular gap function (i.e.~$M = N$), for which the simplified contact laws assume an even cleaner form\footnote{Willis \cite{willis1966} demonstrated that the contact area remains elliptic in the exact solution for a circular gap function and general anisotropic media. Thus, the exact contact law still requires the coupled solution of $e$ and $\phi$ through \eqref{eq:OffsetAngleEccentricityRelations}, while its simplified counterparts simply return a circular contact area as described in Section \ref{sec:SolutionProcedureAnisotropicSpherical}.}. Note that in Section \ref{sec:SolutionProcedureAnisotropicSpherical}, we have described the circular limit in the context of two spherical contacting bodies of radii $R
^{B_1}$ and $R
^{B_2}$, in which case $M = N = 1/2R$ with $1/R = 1/R^{B_1} + 1/R^{B_2}$. From here on, we will refer to the simplified laws described in Sections \ref{sec:Approximation1} and \ref{sec:Approximation2} as the \textit{truncated} and \textit{ad hoc} contact laws, respectively. 

\section{Validation of the contact force laws}
\label{sec:ComparisonContactForceLaws}

\begin{figure}[tb!]
\centering
\includegraphics[width=0.5\textwidth]{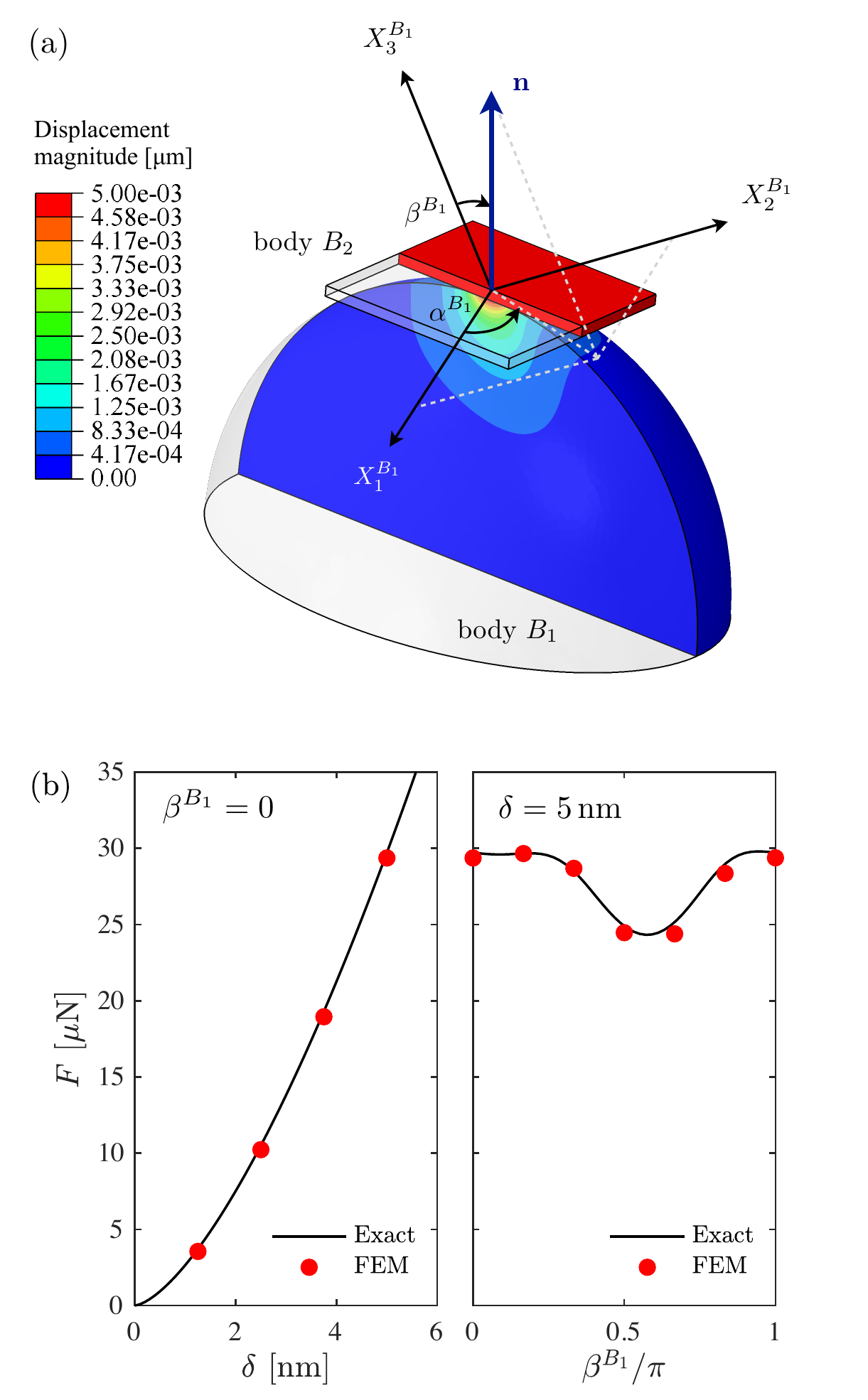}
\caption{(a) A flat rigid plate $B_2$ is assigned a vertical displacement $\delta$ into a smooth elastic body $B_1$ with possibly unequal principal radii of curvature. The orientation of the local coordinate basis of $B_1$ with respect to the contact normal direction $\mathbf{n}$ is parameterized by the Euler angles $\alpha^{B_1}$ and $\beta^{B_1}$. The contour levels on the vertical cut display the magnitude of the elastic displacement generated by an indentation depth $\delta = 5 \, \text{nm}$, as computed in FEM. (b) Contact force predicted by the exact contact law (solid line) and the FEM simulations (red dots) for $\alpha^{B_1} = \pi/2$ and $\beta^{B_1} = 0$ as a function of $\delta$ (left pane), and for $\alpha^{B_1} = \pi/2$ and $\delta = 5 \, \text{nm}$ as a function of $\beta^{B_1}$ (right pane).}
\label{fig:Abaqus}
\end{figure}

In this section, we first validate the accuracy of the exact contact force law against finite-element method (FEM) simulations, before comparing the accuracy of the two simplified contact force laws with respect to their exact counterpart. To this effect, we perform numerical calculations of the force experienced by a flat rigid plate (body $B_2$) pressing against a smooth elastic body made of a given material (body $B_1$), as pictured in Figure \ref{fig:Abaqus}(a).
Following our previous convention, we parameterize the direction of the unit normal $\mathbf{n}$ to the contact plane with respect to the local coordinate basis $(X_1^{B_1},X_2^{B_1},X_3^{B_1})$ of body $B_1$ by the two Euler angles $\alpha^{B_1}$ and $\beta^{B_1}$ depicted in Figure \ref{fig:Abaqus}(a). 

We consider a number of possible scenarios by changing (\textit{i}) the material, (\textit{ii}) the orientation of $B_1$, represented by $\alpha^{B_1}$ and $\beta^{B_1}$, as well as (\textit{iii}) the geometry of the smooth surface, defined by the gap function \eqref{eq:GapFunction}. In terms of materials, we selected three different crystals spanning a wide range of degrees of symmetry. First is iron (Fe), which has a cubic crystalline structure described by  three independent elastic constants $C_{11} = 231$, $C_{44} = 116$, and $C_{12} = 135$ GPa, as determined in \cite{rotter1966}. Second is quartz ($\text{SrO}_2$), which has a trigonal crystalline structure described by 6 independent elastic constants, measured by \cite{heyliger2003} as $C_{11} = 87.2$, $C_{33} = 106$, $C_{44} = 57.2$, $C_{12} = 6.57$, $C_{13} = 12.0$, and $C_{14} = -17.2$ GPa. Finally, third is zirconia ($\text{ZrO}_2$), which has a monoclinic crystalline structure described by 13 independent elastic constants, which were characterized by \cite{chan1991} as $C_{11} = 361$, $C_{22} = 408$, $C_{33} = 258$, $C_{44} = 99.9$, $C_{55} = 81.2$, $C_{66} = 126$, $C_{12} = 142$, $C_{13} = 55.0$, $C_{15} = -21.3$, $C_{23} = 196$, $C_{25} = 31.2$, $C_{35} = -18.2$, and $C_{46} = -22.7$ GPa.

\subsection{Validation of the exact contact force law}

We begin by comparing predictions of the exact contact force law with results from FEM simulations. The 3D setup presented in Figure \ref{fig:Abaqus}(a) is implemented in ABAQUS (2017). The elastic body $B_1$ is designed with principal radii of curvature at the contact point of $R_1 = 0.5 \, \mu\text{m}$ and $R_2 = 0.25 \, \mu\text{m}$, corresponding to gap function coefficients $M = 1/2R_1 = 1 \mu\text{m}^{-1}$ and $N = 1/2R_2 = 2 \mu\text{m}^{-1}$ \cite{barber2018}. The contact interaction between $B_1$ and $B_2$ is modeled following the `surface-to-surface' formulation, with hard contact in the normal direction and no friction in the tangential direction. We assign the elastic properties of quartz ($\text{SrO}_2$) to $B_1$ by specifying its full elasticity tensor and rotating the corresponding material directions $(X_1^{B_1},X_2^{B_1},X_3^{B_1})$ according to the desired values of $\alpha^{B_1}$ and $\beta^{B_1}$. Body $B_1$ is discretized using 112752 quadratic tetrahedral elements (C3D10M), and its base is pinned in the vertical direction. An incremental vertical displacement directed into $B_1$ is prescribed to the flat rigid plate $B_2$, which is defined as an `analytical rigid surface'. The analysis is carried out using the explicit solver by moving the rigid plate at a rate slow enough to ensure that the deformation proceeds in a quasi-static manner, as verified by the fact that (\textit{i}) the total kinetic energy never exceeds 0.03\% of the total internal energy, and (\textit{ii}) the contact force measured at the plate is within 1\% equal to the sum of the vertical reaction forces at the basal nodes of $B_1$.

The contour levels on the vertical cut of body $B_1$ in Figure \ref{fig:Abaqus}(a) display the magnitude of the elastic displacements induced by an indentation depth $\delta = 5 \, \text{nm}$, as computed in FEM, for material orientation $\alpha^{B_1} = \beta^{B_1} = \pi/2$. Interestingly, the anisotropy of the constitutive relation is reflected in the absence of axisymmetry (with respect to the contact normal direction) of the elastic displacement field. For a more quantitative analysis, the red dots in the left pane of Figure \ref{fig:Abaqus}(b) depict the FEM contact force for material orientation $\alpha^{B_1} = \pi/2$, $\beta^{B_1} = 0$ and four different values of the vertical displacement $\delta$ of $B_2$ into $B_1$. These FEM results are in excellent agreement with the corresponding predictions from the exact contact force law shown by the solid line; the power $3/2$-dependence of $F$ on $\delta$ is also clearly visible. Conversely, the red dots in the right pane of Figure \ref{fig:Abaqus}(b) display the FEM contact force for $\delta = 5 \, \text{nm}$ and different material orientations defined by $\alpha^{B_1} = \pi/2$ and varying values of $\beta^{B_1}$. Once again, the FEM results agree well with the exact contact force law shown by the solid line, with the difference between the two not exceeding 3.4\%.

The slight discrepancy between FEM and theoretical results observed in Figure \ref{fig:Abaqus}(b) can be attributed to various reasons. On the one hand, the FEM solution is dependent on the resolution of the mesh in the vicinity of the contact region, and further refinement of the mesh would reduce errors arising from the numerical discretization. On the other hand, the exact contact force law relies on the assumptions that the size of the contact area is small with respect to the dimensions of $B_1$ as well as its radii of curvature at the contact point. Such assumptions are never satisfied exactly, thus invariably lead to small errors when the contact law is applied to a real-case scenario. Notwithstanding, the comparisons displayed in Figure \ref{fig:Abaqus}(b) exhibit a sufficient level of agreement to validate both the accuracy and the implementation of the exact contact force law.

\subsection{Accuracy of the simplified contact force laws}

\subsubsection{Polar visualizations}
\label{sec:PolarVisualizations}

We now proceed with the comparison of the two simplified contact force laws with respect to their exact counterpart. As in the previous section, we calculate the contact force experienced in the setup pictured in Figure \ref{fig:Abaqus}(a), this time using a wider range of materials, orientations and surface geometries. We first show polar visualizations of the directional dependence of the force predicted by the exact solution and its two simplifications for an indentation depth (overlap) $\delta = 100 \, \text{nm}$. Due to the symmetry exhibited by the Green's function \eqref{eq:GreenFunctionAnisotropic} with respect to the sign of the unit normal $\mathbf{n}$, the behavior of the force is completely specified for all materials by the hemisphere $\alpha^{B_1} \in [0,2\pi]$, $\beta^{B_1} \in [0, \pi/2]$. Thus, we visualize the directional dependence of the contact force by projecting each direction point on the hemisphere to a plane through stereographic projection, in such a way that the data pertaining to the orientation $(\alpha^{B_1}, \beta^{B_1})$ will be displayed at the location $(\tan(\beta^{B_1}/2) \cos \alpha^{B_1}, \tan(\beta^{B_1}/2) \sin \alpha^{B_1})$ in a disk of unit radius. 

\begin{figure*}
\centering
\includegraphics[width=\textwidth]{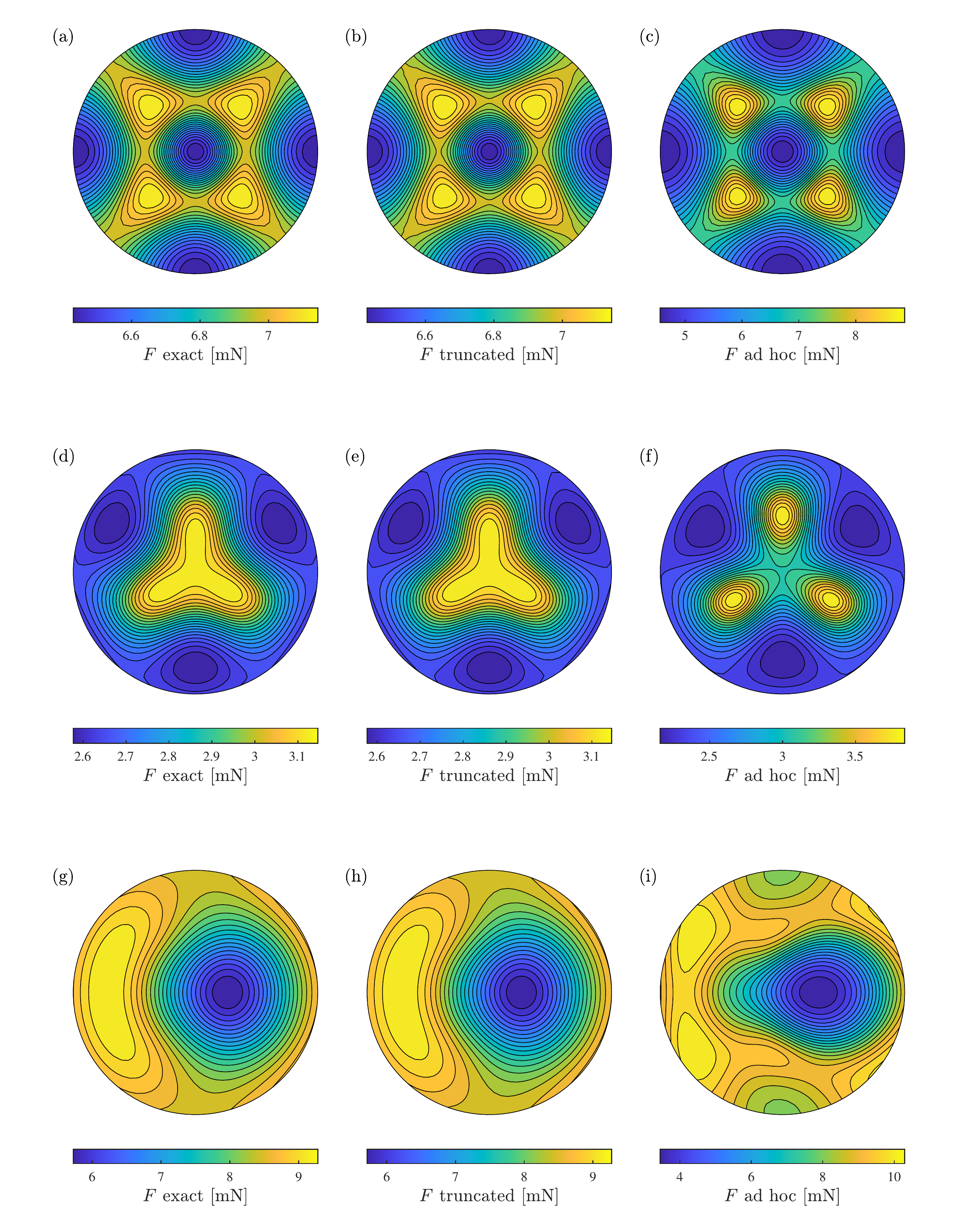}
\caption{Polar visualizations of the normal force $F$ predicted by the exact contact force law and its two simplifications for iron (a,b,c), quartz (d,e,f), and zirconia (g,h,i), under indentation depth $\delta = 100 \, \text{nm}$ and gap function coefficients $M = N = 1 \, \mu\text{m}^{-1}$.}
\label{fig:ForceComparison_BoA1}
\end{figure*}

\begin{figure*}
\centering
\includegraphics[width=\textwidth]{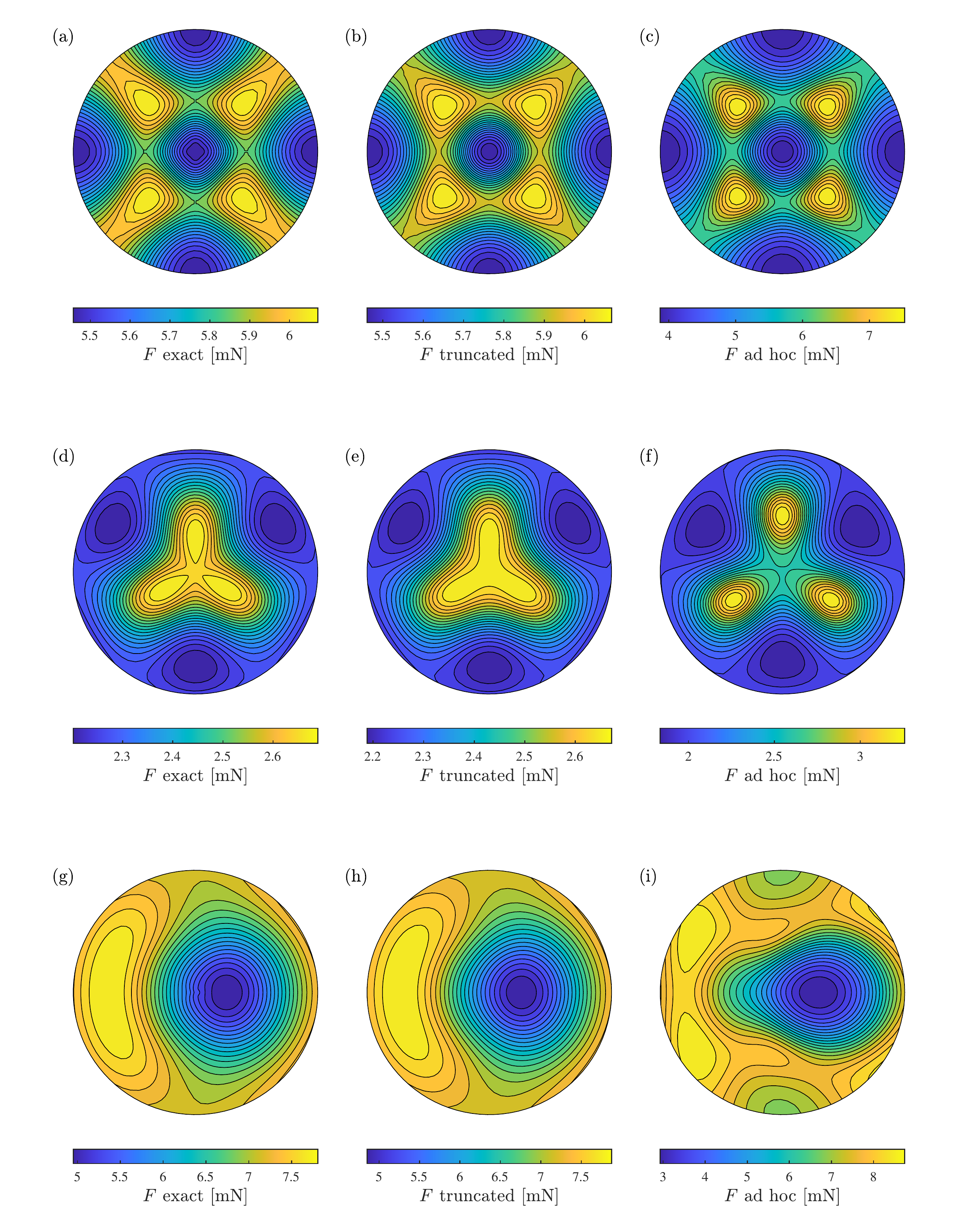}
\caption{Polar visualizations of the normal force $F$ predicted by the exact contact force law and its two simplifications for iron (a,b,c), quartz (d,e,f), and zirconia (g,h,i), under indentation depth $\delta = 100 \, \text{nm}$ and gap function coefficients $M = 1 \, \mu\text{m}^{-1}$ and $N = 2 \, \mu\text{m}^{-1}$.}
\label{fig:ForceComparison_BoA2}
\end{figure*}

Figures \ref{fig:ForceComparison_BoA1} and \ref{fig:ForceComparison_BoA2} display such polar visualizations of the exact contact force law and its two simplifications for a circular gap function ($M = N = 1 \, \mu\text{m}^{-1}$) in Figure \ref{fig:ForceComparison_BoA1} and an elliptic gap function ($M = 1 \, \mu\text{m}^{-1}$, $N = 2 \, \mu\text{m}^{-1}$) in Figure \ref{fig:ForceComparison_BoA2}.
In the elliptic case, we have chosen to orient the principal axes $(x,y)$ of the gap function along the polar $(\alpha, \beta)$ directions. For both figures, (a,b,c) correspond to iron, (d,e,f) to quartz, and (g,h,i) to zirconia. 
Surprisingly, we notice that the truncated force law is remarkably close to the exact solution for all materials, contact directions and shapes of the gap function. This result is extremely promising for DEM applications since the truncated law can return a near-exact contact force at a very reasonable cost (presuming that one uses a look-up table approach as described in Section \ref{sec:Approximation1} and Appendix \ref{app:CalculationCompositePSM}). The ad hoc approximation, on the other hand, deviates further away from the exact solution. The accuracy with which it predicts the shape of the contour levels of the force depends on the degree of symmetry of the material -- it performs very well in this regard for iron, reasonably well for quartz, and more poorly for zirconia. More importantly, it fails to correctly predict the extrema of the force and displays a much stronger dependence of the latter on the contact normal direction, as compared with the other solutions.

The relative dependence of the force on the contact normal direction is not strongly influenced by the geometry of the gap function, as one observes by comparing Figures \ref{fig:ForceComparison_BoA1} and \ref{fig:ForceComparison_BoA2}. The most apparent difference between the circular and elliptic gap functions under a given overlap distance is that the force is lower in the elliptic case for all contact normal orientations, which is expected since the case $M = 1 \, \mu\text{m}^{-1}$, $N = 2 \, \mu\text{m}^{-1}$ has a higher mean curvature at the contact point than the case $M = N = 1 \, \mu\text{m}^{-1}$. Curiously, the other geometrical features of the exact solution -- namely the eccentricity $e$ and orientation $\phi$ of the contact area -- are more sensitive to the geometry of the gap function, as shown in Appendix \ref{app:GeometricFeaturesExactHertzianSolution}.

\subsubsection{Error analysis}

\begin{figure}[tb!]
\centering
\includegraphics[width=0.5\textwidth]{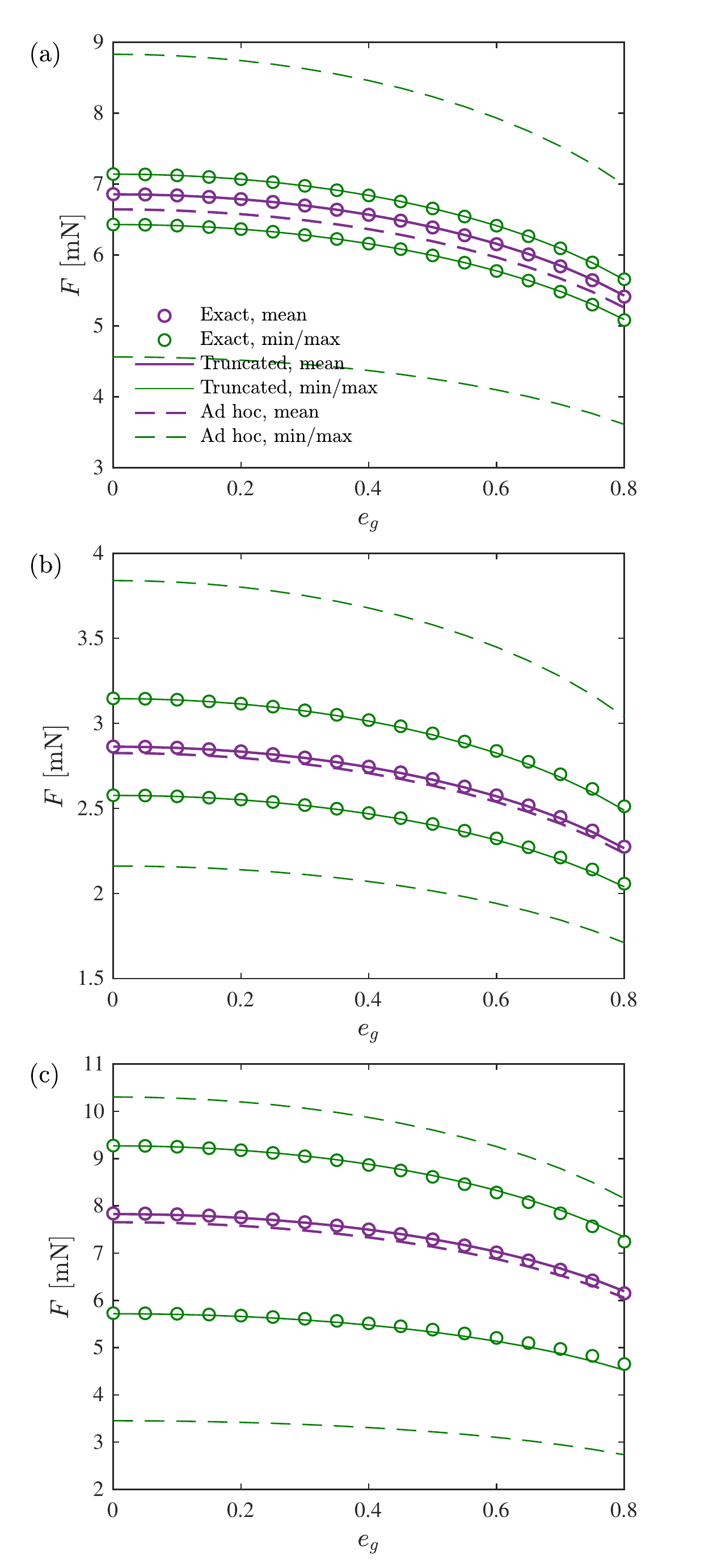}
\caption{Mean, maximum and minimum values of the normal force $F$, over all contact normal directions, predicted by the exact and simplified contact laws as a function of the eccentricity $e_g$ of the gap function, for (a) iron, (b) quartz, and (c) zirconia.}
\label{fig:BoA_error}
\end{figure}

\begin{figure}[tb!]
\centering
\includegraphics[width=0.5\textwidth]{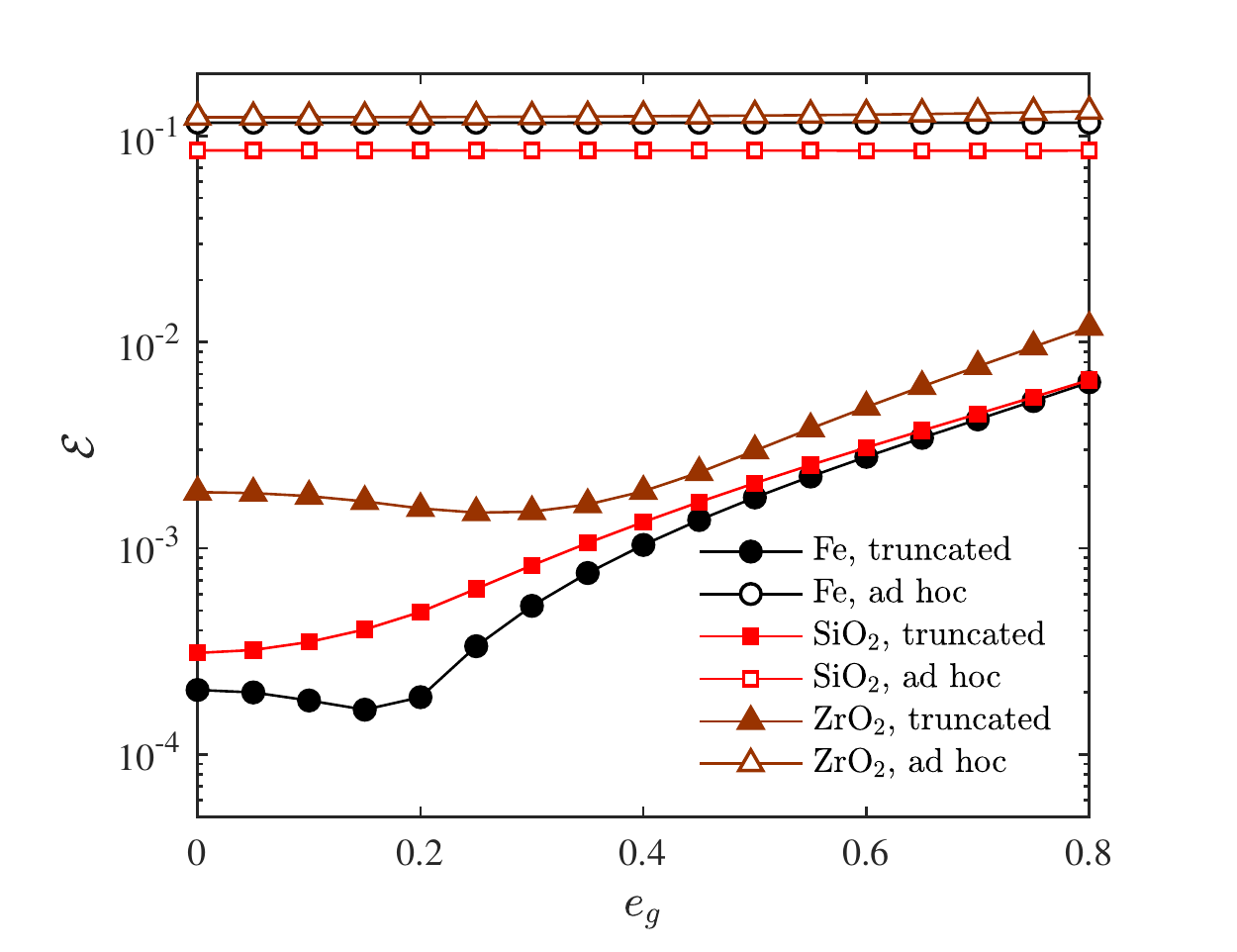}
\caption{Mean relative error $\mathcal{E}$ over all contact normal directions of the force $F$ predicted by the two simplified contact laws as a function of the eccentricity $e_g$ of the gap function, for iron, quartz, and zirconia.}
\label{fig:BoA_error_avg}
\end{figure}

Next, we perform a quantitative analysis of the accuracy of the two simplified contact force laws with respect to their exact counterpart. Let us first define $e_g$, the eccentricity of the gap function, as
\begin{equation}
e_g = \sqrt{1-\frac{M}{N}}.
\end{equation}
The quantity $e_g$ measures the eccentricity of the contour levels of the gap function $g_0(x,y)$, in the same way that $e$ quantifies the eccentricity of the boundary of the contact area. A circular gap function corresponds to $e_g = 0$. We first quantify the sensitivity of the various contact laws with respect to the contact normal direction for different values of $e_g$, using the same  overlap $\delta = 100 \, \text{nm}$ as prescribed before. To this effect, Figure \ref{fig:BoA_error} shows the mean and extrema values, over all contact normal directions, of the normal force predicted by the exact and simplified contact laws versus $e_g$, for iron (a), quartz (b) and zirconia (c).
 We again observe that the truncated law is very accurate, while the ad hoc law exaggerates the dependence of the force on the orientation. For a more quantitative comparison, we define, for a given value of $e_g$ and a given material, the relative error
\begin{equation}
\mathcal{E} = \frac{1}{2\pi} \int_0^{2\pi} \int_0^{\pi/2} \frac{|F_\text{s}(\alpha,\beta)-F_\text{e}(\alpha,\beta)|}{F_\text{e}(\alpha,\beta)} \sin \beta d\alpha d\beta,
\label{eq:MeanRelativeError}
\end{equation}
where $F_\text{e}$ and $F_\text{s}$ refer, respectively, to the exact and simplified solutions. Thus, \eqref{eq:MeanRelativeError} returns the mean relative error over all contact normal orientations. Figure \ref{fig:BoA_error_avg} shows the error as a function of the eccentricity of the gap function $e_g$ for the three materials considered previously.
The accuracy of the truncated law is remarkable for small values of the gap function eccentricity $e_g$, and remains very good as $e_g$ increases, with the relative error $\mathcal{E}$ remaining near or under 1\%. The ad hoc law, on the other hand, behaves more poorly with the error being on the order of 10\% for the three materials.

\begin{figure}
\centering
\includegraphics[width=0.5\textwidth]{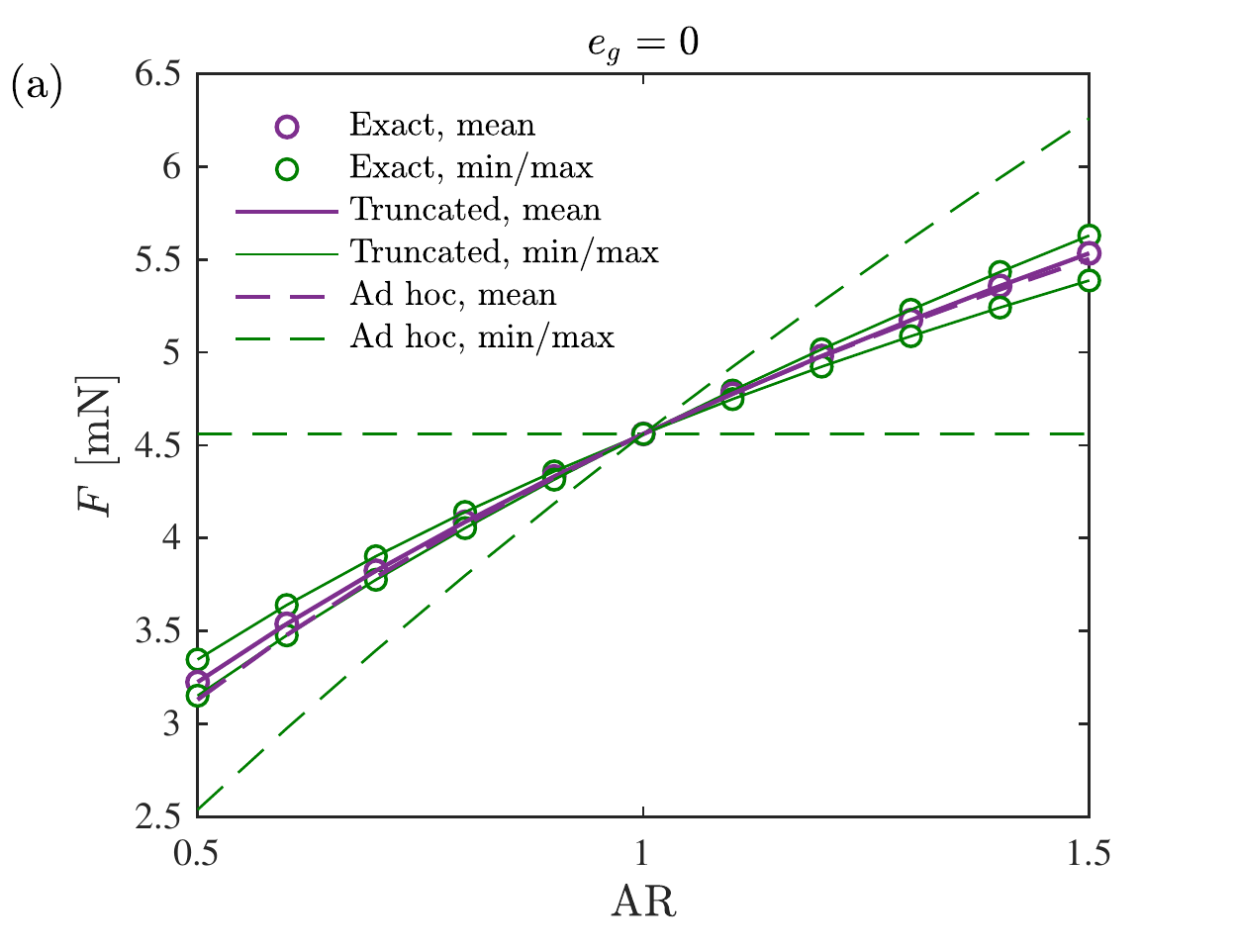}
\caption{(a) Mean, maximum and minimum values of the normal force $F$, over all contact normal directions, predicted by the exact and simplified contact laws for an arbitrary cubic material with varying anisotropy ratio $\text{AR}$ and $e_g = 0$.}
\label{fig:AR_error}
\end{figure}

\begin{figure}
\centering
\includegraphics[width=0.5\textwidth]{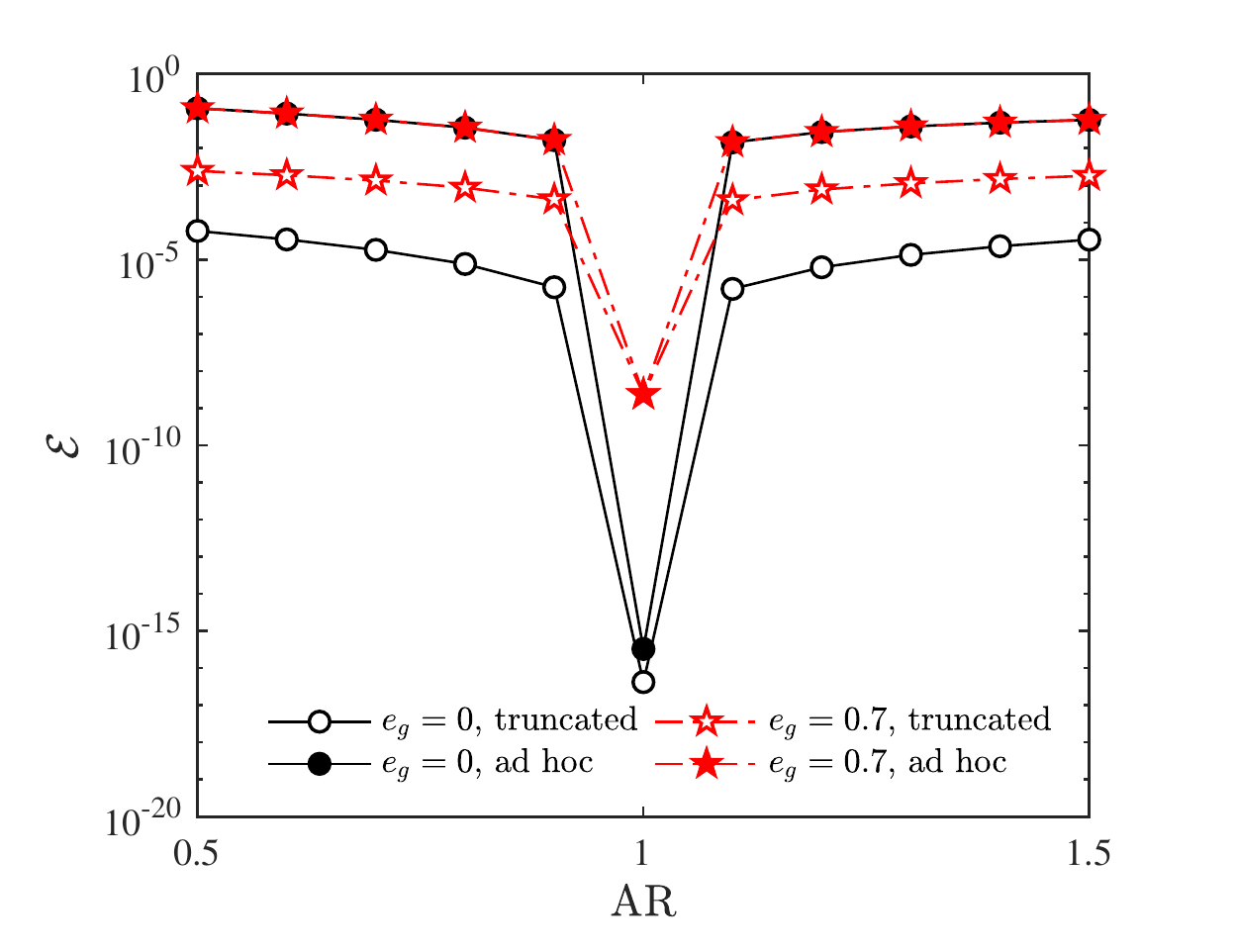}
\caption{Mean relative error $\mathcal{E}$ over all contact normal directions of the force $F$ predicted by the two simplified contact laws as a function of the anisotropy ratio $\text{AR}$ of an arbitrary cubic material.}
\label{fig:AR_error_avg}
\end{figure}

We now study the behavior of the exact and simplified contact laws as the constitutive relation approaches the isotropic limit. For this purpose, we construct an arbitrary cubic material of varying anisotropy ratio $\text{AR}$ defined by \cite{zener1948} as
\begin{equation}
\text{AR} = \frac{2C_{44}}{C_{11}-C_{12}},
\end{equation}
with the particular case $\text{AR} = 1$ corresponding to an isotropic material, and we pick the same values for $C_{11}$ and $C_{12}$ as for iron. (For reference, iron then corresponds to the case $\text{AR} = 2.41$.) Figure \ref{fig:AR_error} shows the mean and extrema values, over all contact normal directions, of the normal force predicted by the exact and simplified contact laws for this arbitrary material and a circular gap function, that is, $e_g = 0$. 
As expected, the two simplified contact laws degenerate to the exact solution in the limiting case $\text{AR} = 1$ of an isotropic material. This remains true for a finite value of $e_g$, as displayed in Figure \ref{fig:AR_error_avg} for $e_g = 0.7$ in terms of the mean relative error $\epsilon$ defined in \eqref{eq:MeanRelativeError}.
Here again, the truncated contact law is remarkably close to the exact solution for all values of anisotropy ratio $\text{AR}$ and gap function eccentricity $e_g$, with the relative error $\mathcal{E}$ remaining under 1\%, while the second approximation returns larger relative errors on the order of 10\%. 

Going forward, we select the truncated simplification as the contact force law of choice for implementation into a DEM code. Implemented with a look-up table approach, the computation of this force law is equally fast as the ad hoc simplification, yet returns results that are accurate to within 1\% for ellipsoids and within 0.1\% for spherical particles.

\section{Applications}
\label{sec:Applications}

We now implement the truncated anisotropic contact law described in Section \ref{sec:Approximation1} into a custom DEM code, which enables the simulation of granular materials composed of elastically anisotropic particles. Even though the contact laws that we have derived are applicable to arbitrarily-shaped bodies as long as their surfaces are smooth and convex, we will here restrict ourselves to spherical particles. We will show two examples, one static and one dynamic, where the anisotropy of the constitutive relation induces changes in the macroscopic properties of the system.

\subsection{Equations of motion}
\label{sec:EquationsMotion}

Consider a system of $N$ spherical particles $i = 1, \dots, N$ with (possibly distinct) elasticity tensors $\mathbb{C}_{i}$. In this section, we adopt a slight change of notation and indicate quantities pertaining to body $i$ with a subscript $i$, in line with conventions from the DEM literature. The positions and orientations of the particles are described by a set of generalized coordinates $\{\mathbf{q}_i\} = (\{\mathbf{r}_i\},\{\bm{\epsilon}_i\})$, where $\{\mathbf{r}_i\} \in \mathbb{R}^3$ denotes the position of the center of mass of body $i$, and $\{\bm{\epsilon}_i\} \in \mathbb{R}^4$ is a set of Euler parameters (unit quaternions) that characterizes the orientation of body $i$, both in the global reference frame $(\mathcal{X}_1,\mathcal{X}_2,\mathcal{X}_3)$ shown in Figure \ref{fig:GeometryBody}(a). The linear and angular velocities of the particles are described by generalized velocities $\{\mathbf{v}_i\} = (\{\dot{\mathbf{r}}_i\},\{\bm{\omega}_i\})$, where $\{\bm{\omega}_i\} \in \mathbb{R}^3$ is the angular velocity of body $i$ in the global frame and relates to the time derivative of the Euler parameters $\{\bm{\epsilon}_i\}$ as (see \cite{evans1977,dvziugys2001} or equation (9.3.37) in \cite{haug1989})
\begin{equation}
\{\dot{\bm{\epsilon}_i}\} = \frac{1}{2} [\mathbf{A}(\bm{\epsilon}_i)] \{\bm{\omega}_i\},
\label{eq:EvolutionEulerParameters}
\end{equation}
with the matrix $[\mathbf{A}(\bm{\epsilon}_i)] \in \mathbb{R}^{4 \times 3}$ defined as
\begin{equation}
[\mathbf{A}(\bm{\epsilon}_i)] = \left[
\begin{array}{ccc}
-\epsilon_{i,1} & -\epsilon_{i,2} & -\epsilon_{i,3} \\
\epsilon_{i,0} & \epsilon_{i,3} & -\epsilon_{i,2} \\
-\epsilon_{i,3} & \epsilon_{i,0} & \epsilon_{i,1} \\
\epsilon_{i,2} & -\epsilon_{i,1} & \epsilon_{i,0}
\end{array}
\right].
\end{equation} 
Two bodies $i$ and $j$, with diameters $d_i$ and $d_j$, interact when their signed overlap function,
\begin{equation}
\delta_{ij} = \frac{d_i + d_j}{2} - |\mathbf{r}_i - \mathbf{r}_j|,
\end{equation}
is positive. Denoting by $c_i = \{j : \delta_{ij} \ge 0 \}$ the set of particles that are in contact with body $i$, the generalized velocities can be integrated in time using Newton's equations of motion,
\begin{subequations}
\begin{align}
m_i \{\ddot{\mathbf{r}}_i\} &= \sum_{j \in c_i} \{\mathbf{F}_{ij}\} + m_i \{\mathbf{g}\}, \\
I_i \{\dot{\bm{\omega}}_i\} &= \sum_{j \in c_i} (a_{ij} \{\mathbf{n}_{ij}\} \times \{\mathbf{F}_{ij}\}),
\end{align}
\label{eq:EvolutionVelocities}%
\end{subequations}
where $m_i$ and $I_i$ denote respectively the mass and moment of inertia of particle $i$. At each contact, $a_{ij} = (d_i-\delta_{ij})/2$ denotes the distance from the center of mass of particle $i$ to its contact point with particle $j$, the unit normal vector $\mathbf{n}_{ij} = (\mathbf{r}_j-\mathbf{r}_i)/|\mathbf{r}_j-\mathbf{r}_i|$ is directed from $i$ to $j$, and the force $\{\mathbf{F}_{ij}\}$ consists of normal and tangential components,
\begin{equation}
\{\mathbf{F}_{ij}\} = F_{ij}^n \{\mathbf{n}_{ij}\} + F_{ij}^t \{\mathbf{t}_{ij}\},
\end{equation}
where the tangent unit vector $\mathbf{t}_{ij}$ belongs to the contact plane and depends on the history of relative tangential velocities of $i$ and $j$ at the contact point. In this paper, we consider frictionless\footnote{A direct consequence of this assumption is that in the absence of external torques, spherical particles will keep their initial orientation throughout the simulation. Nevertheless, our exposition accounts for the possible presence of angular velocities in an effort to be as general as possible. Torques may arise in other works as a result of the geometry or surface roughness of the particles, and it is critical to treat their orientations correctly since the anisotropic contact law is orientation-dependent.} bodies so that $F_{ij}^t = 0$. The normal force $F_{ij}^n$ comprises an elastic and a dissipative part,
\begin{equation}
F_{ij}^n = -\max(F_{ij}^e + F_{ij}^d,0),
\end{equation}
where the $\max(\cdot)$ function forbids the existence of a cohesion force, and the orientation-dependent elastic component $F_{ij}^e$ is given by the normal contact force law derived in Section \ref{sec:Approximation1},
\begin{equation}
F_{ij}^e = \frac{4}{3} \tilde{E}_*^c(\{\bm{\epsilon}_i\},\{\bm{\epsilon}_j\},\{\mathbf{n}_{ij}\}) R_{ij}^{1/2} \delta_{ij}^{3/2},
\label{eq:ContactLawDEM}
\end{equation}
where $1/R_{ij} = (2/d_i+2/d_j)$. Likewise, one expects the dissipative component $F_{ij}^d$ to inherit an orientation dependence from the anisotropy of the material structure. However, deriving such a relation falls outside the scope of this paper, and we restrict ourselves to the standard isotropic expression
\begin{equation}
F_{ij}^d = \gamma_n \dot{\delta}_{ij},
\end{equation}
where $\gamma_n$ is a constant damping coefficient. 
This simplification is reasonable for flowing granular materials, where damping is known to play a negligible role within a particular range of strain rates \cite{da2005}. For quasi-static problems, the form of the dissipation is inconsequential so long as one is interested in static quantities after particles have come to a rest, which is the case of our only upcoming example using a nonzero damping coefficient $\gamma_n$.

In equation \eqref{eq:ContactLawDEM}, the dependence of $\tilde{E}_*^c$ on the relative orientations of bodies $i$ and $j$ with respect to the contact normal direction has been indicated through the Euler parameters $\{\bm{\epsilon}_i\}$, $\{\bm{\epsilon}_j\}$ and the components $\{\mathbf{n}_{ij}\}$ of the contact normal, which are readily available in the simulation. Given these inputs, we present in Appendix \ref{app:CalculationCompositePSM} an algorithm to retrieve the value of $\tilde{E}_*^c$ from two (or one, if $\mathbb{C}_{i} = \mathbb{C}_{j}$) precomputed tables of values of the plane strain modulus, $[\tilde{E}_*](\cdot,\cdot\,;\mathbb{C}_{i})$ and $[\tilde{E}_*](\cdot,\cdot\,;\mathbb{C}_{j})$, the computation of which is described in Appendix \ref{app:CalculationPSM}. These algorithms for the calculation of the look-up tables and contact force have been provided as a Python code in an online repository at \url{https://github.com/smowlavi/AnisotropicGrains.git}. Further details regarding the numerical implementation of the DEM code and parameter values are listed in Appendix \ref{app:DetailsDEM}.

\subsection{Static force distribution in a pyramid}
\label{sec:StaticForceDistributionPyramid}

As a first example, we consider a static square-based pyramid of close-packed single-crystal zirconia spheres, with ten particles along each side of the base. This system is statically indeterminate due to each interior particle possessing twelve neighbors \cite{olsen2018}. Therefore, the equilibrium contact forces will depend on the contact stiffnesses (that is, on the composite plain strain moduli $\tilde{E}_*^c$), which in the case of anisotropic particles are a function of the contact directions and particle orientations. As we will show next, our anisotropic DEM framework enables us to investigate the relationship between the floor pressure at the base of the pyramid and the orientation of the particles.

Notice from Figure \ref{fig:ForceComparison_BoA1} that a sphere made of single-crystal zirconia may be thought of as having a band of high contact stiffness along its equator. We will consider four separate arrangements in which every particle is either oriented such that the strong band is roughly horizontal (orientation 1), resulting in all contacts witnessing approximately the same stiffness from that particle; or the strong band is roughly aligned with a vertical plane parallel to the $y = x$ diagonal of the square base (orientation 2), causing stiffer contacts oriented in those directions compared to those oriented in the other direction.

\begin{figure}[tb!]
\centering
\includegraphics[width=0.5\textwidth]{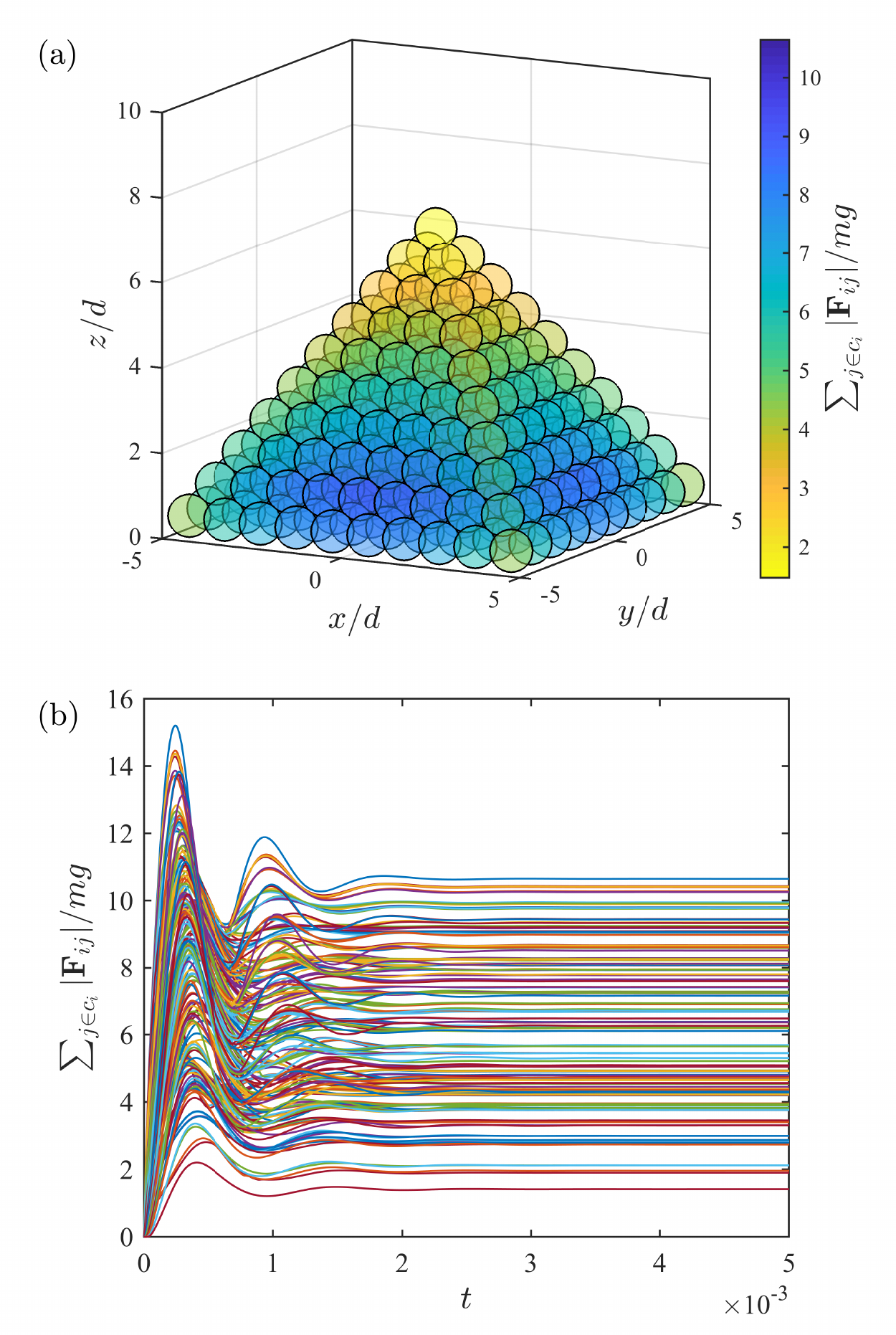}
\caption{(a) Geometry of the static square-based pyramid. Each particle $i$ is colored according to $\sum_{j\in c_i} |\mathbf{F}_{ij}|/mg$, the normalized sum of the force magnitudes that it withstands at all its contact points. (b) The same quantity is plotted over time during the settling of the pyramid.}
\label{fig:Pyramid}
\end{figure}

Figure \ref{fig:Pyramid}(a) shows a three-dimensional visualization of the pyramid, which is initialized by placing the particles in a position where they barely touch their neighbors.
The pyramid is then allowed to settle under the acceleration of gravity, with the contact forces oscillating during a transient phase before reaching their equilibrium values. For the case of all particles following orientation 1, Figure \ref{fig:Pyramid}(b) shows this phenomenon through the time evolution of $\sum_{j\in c_i} |\mathbf{F}_{ij}|/mg$, the normalized sum of the force magnitudes that each particle experiences at all its contact points. The same quantity is shown at final time in Figure \ref{fig:Pyramid}(a) as the semi-transparent color applied to each grain.

\begin{figure*}[tb!]
\centering
\includegraphics[width=\textwidth]{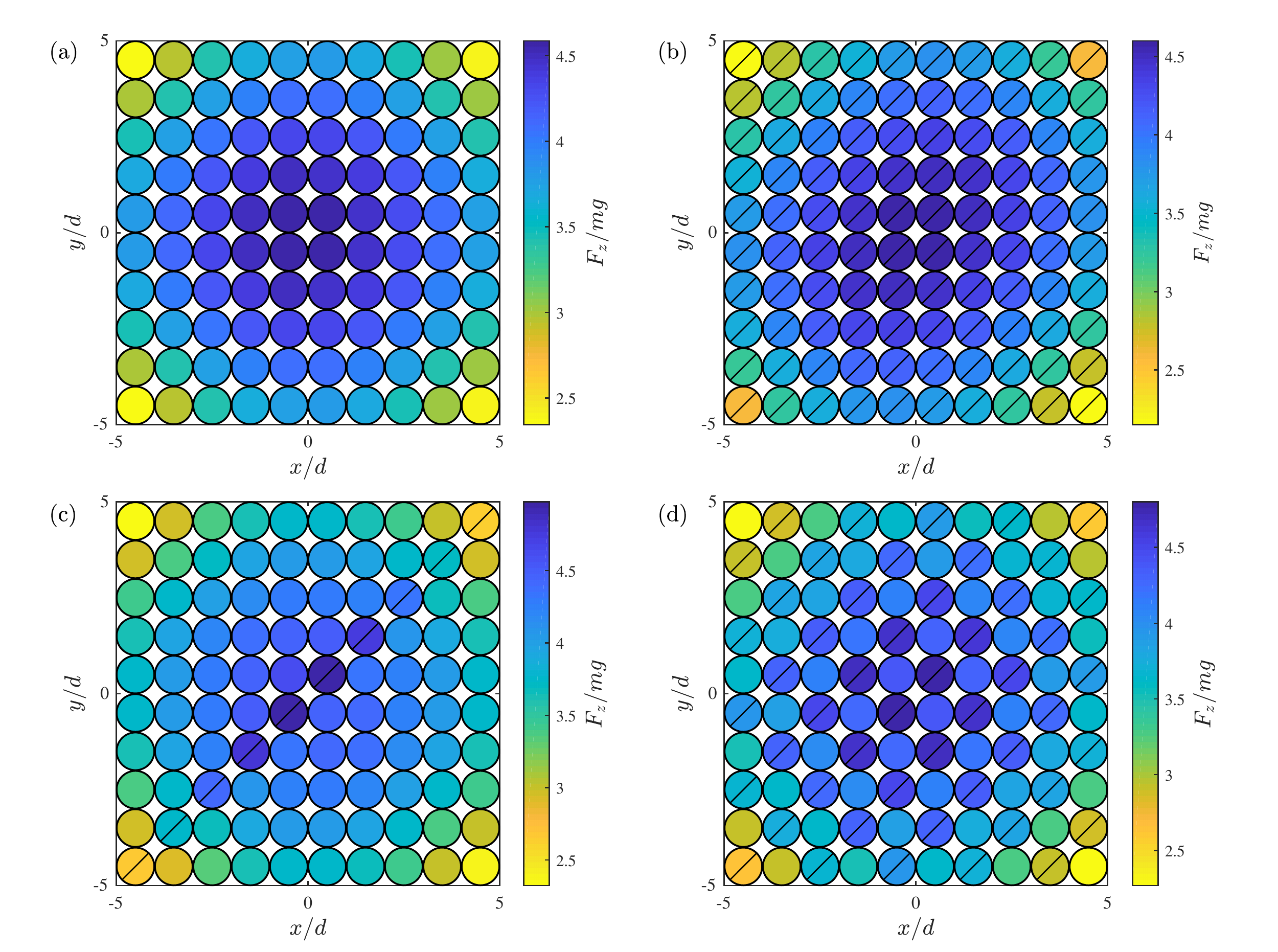}
\caption{Distribution of normalized reaction forces $F_z/mg$ on the base of the pyramid for particles oriented such that either all contacts see uniform stiffness from the particle, or contacts along directions parallel to the $y = x$ plane are stronger. Particles belonging to the second group are displayed with a diagonal line, and particles in upper layers are oriented identically to the base layer particles belonging to the same $y = x+c$ vertical plane.}
\label{fig:PyramidForceBase}
\end{figure*}

In Figure \ref{fig:PyramidForceBase}, we display the distribution of normalized reaction forces $F_z/mg$ on the base of the pyramid, once equilibrium is reached.
We consider four separate arrangements of particle orientation, which we visualize by displaying particles in orientation 2 with a diagonal line aligned along their strong band. Particles in upper layers are oriented identically to the base layer particles belonging to the same $y = x+c$ vertical plane. As expected, the reaction forces are symmetrical in the case shown in Figure \ref{fig:PyramidForceBase}(a) where all particles have orientation 1, since all the contacts see approximately the same stiffness. That symmetry is broken and a clear effect of anisotropy emerges in Figure \ref{fig:PyramidForceBase}(b), where all particles have orientation 2. Due to the stronger contacts along directions parallel to the $y = x$ plane, the two corner particles aligned along the `strong' $y = x$ diagonal inherit a larger reaction force than the other two corner particles. The picture gets even more interesting in Figures \ref{fig:PyramidForceBase}(c) and (d), which demonstrate that it is possible to tune the reaction force beneath the pyramid by mere rotation of the constituent particles. To conclude, this simple example highlights the importance of accounting for anisotropic effects in the discrete element modeling of elastically anisotropic particles, even in situations that involve no dynamics at all.

\subsection{Sound transmission in a granular chain}
\label{sec:SoundTransmissionGranularChain}

As a second example, we investigate the transmission of sound in a compressed chain of adjacent spherical particles between two fixed walls. A large body of work has researched the behavior and frequency response of such `granular crystals' to small-amplitude dynamic displacements of the particles, where small is in comparison with the static overlap imposed between adjacent particles by the compression force. In particular, different authors have shown that by combining particles with different geometrical or material properties, it is possible to obtain a frequency response characterized by acoustic band gaps inside of which no frequencies are allowed \cite{hladky2005,herbold2009,boechler2011}, thus filtering out input frequencies. Such filters are desirable for a range of purposes ranging from acoustic filters to vibrational isolation, and the tunability of these band gaps is key to delivering optimal performance.

The existence of acoustic band gaps requires the contact stiffnesses between the grains to be non-uniform \cite{jensen2003}. This is most simply achieved in diatomic chains consisting of particles with alternating properties, for which a single band gap appears \cite{brillouin1953,kittel1976,hladky2005}. A second band gap was shown in \cite{boechler2011} to emerge in diatomic chains composed of three-particle unit cells. The tunability of these band gaps requires a change in the properties of the particles, which is typically done by altering their size, geometry or constituent material. Clearly, this is not feasible in practice when one desires to control the band gap frequencies in real-time.

\begin{figure}[tb!]
\centering
\includegraphics[width=0.5\textwidth]{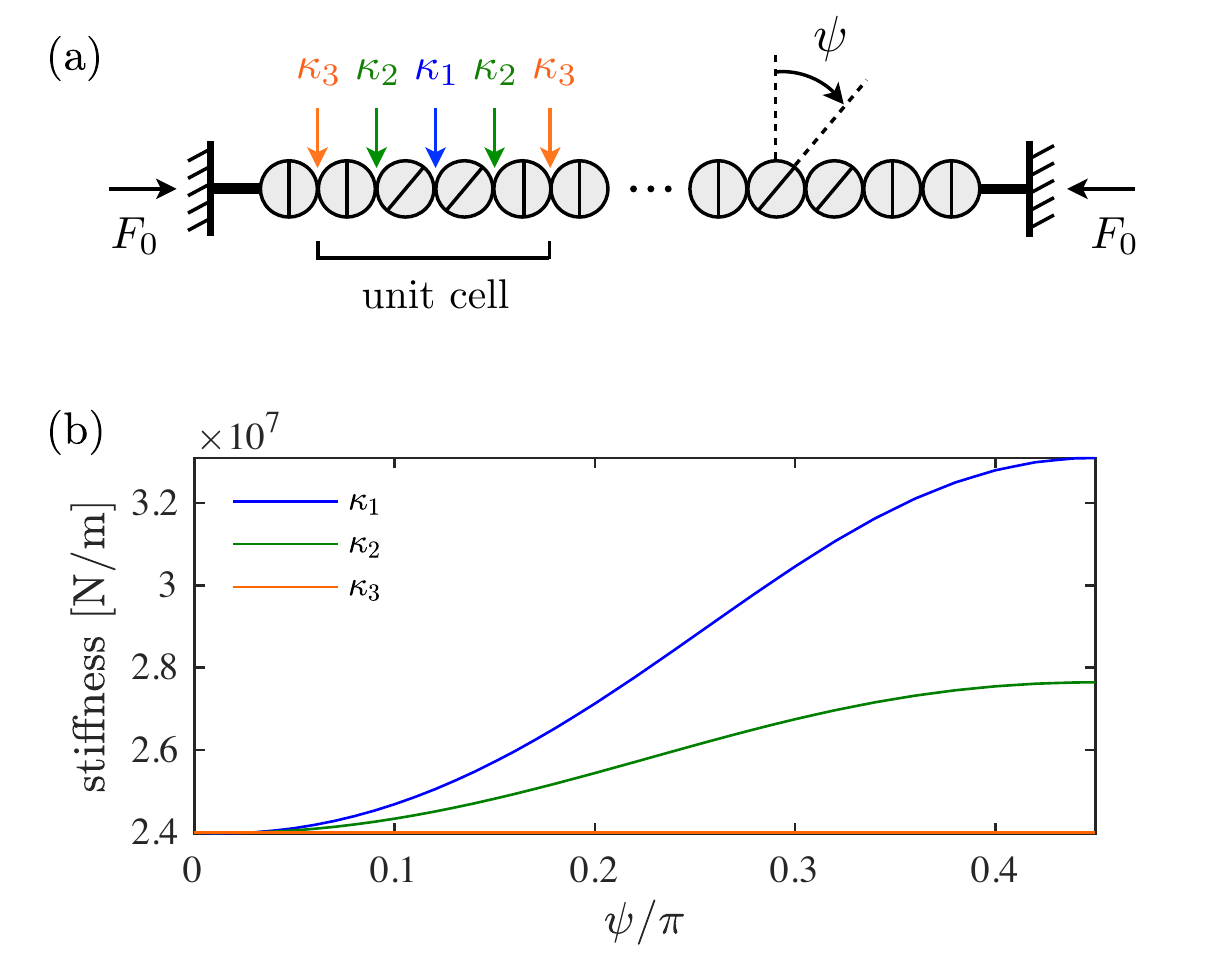}
\caption{(a) Schematic of the granular chain, composed of periodically repeated four-particle unit cells and compressed between two fixed walls. The orientation $\psi$ of the middle two particles in each unit cell is varied systematically, while that of the two edge particles is fixed. (b) Dependence of the linearized contact stiffnesses $\kappa_1$, $\kappa_2$, $\kappa_3$ on the orientation $\psi$ of the middle particles.}
\label{fig:ChainGeometry}
\end{figure}

As we have seen throughout this paper, elastically anisotropic bodies exhibit an orientation-dependent contact stiffness. Here, we utilize this property to construct a monoatomic granular crystal that possesses band gaps that may be tuned by mere rotation of its constituent particles. Specifically, consider the chain of anisotropic zirconia particles pictured in Figure \ref{fig:ChainGeometry}(a) and compressed between two fixed walls.
The chain consists of periodically repeated four-particle unit cells in which the orientation $\psi$ of the middle two particles is varied systematically while that of the two edge particles is kept fixed. The angle $\psi$ is defined as the orientation of the strong band of the zirconia spheres (schematized in Figure \ref{fig:ChainGeometry}(a) by the straight line within each sphere) with respect to the plane orthogonal to the chain axis. The edge particles within each unit cell are oriented such that the strong band is orthogonal to the chain axis.

In the linear regime that we investigate, the relative displacement between any two adjacent particles is small with respect to their static overlap $\delta_{ij}^0$ caused by the compression force $F_0$. Thus, the overlap term $\delta_{ij}^{3/2}$ occurring in the contact force law \eqref{eq:ContactLawDEM} can be linearized about $\delta_{ij}^0$, producing a force-displacement relation that is linear with a proportionality constant termed the linearized contact stiffness. The latter is clearly a function of the composite plain strain modulus $\tilde{E}_*^c$ and therefore depends on the orientation of the particles. (For more details, the reader is invited to refer to Appendix \ref{app:TheoryCompressedChain}.) As pictured in Figure \ref{fig:ChainGeometry}(a), the structure of the unit cell in our granular chain gives rise to three different linearized contact stiffnesses $\kappa_1$, $\kappa_2$, and $\kappa_3$, which depend on the orientation $\psi$ of the middle particles according to Figure \ref{fig:ChainGeometry}(b). Note that $\kappa_3$ is constant since it measures the stiffness between the edge particles of two adjacent unit cells, the orientations of which are fixed. Finally, we neglect dissipation effects, which in practice result in a small uniform shift of the band-gap frequencies but do not change their overall topological features \cite{jensen2003,boechler2010}.

\begin{figure}[tb!]
\centering
\includegraphics[width=0.5\textwidth]{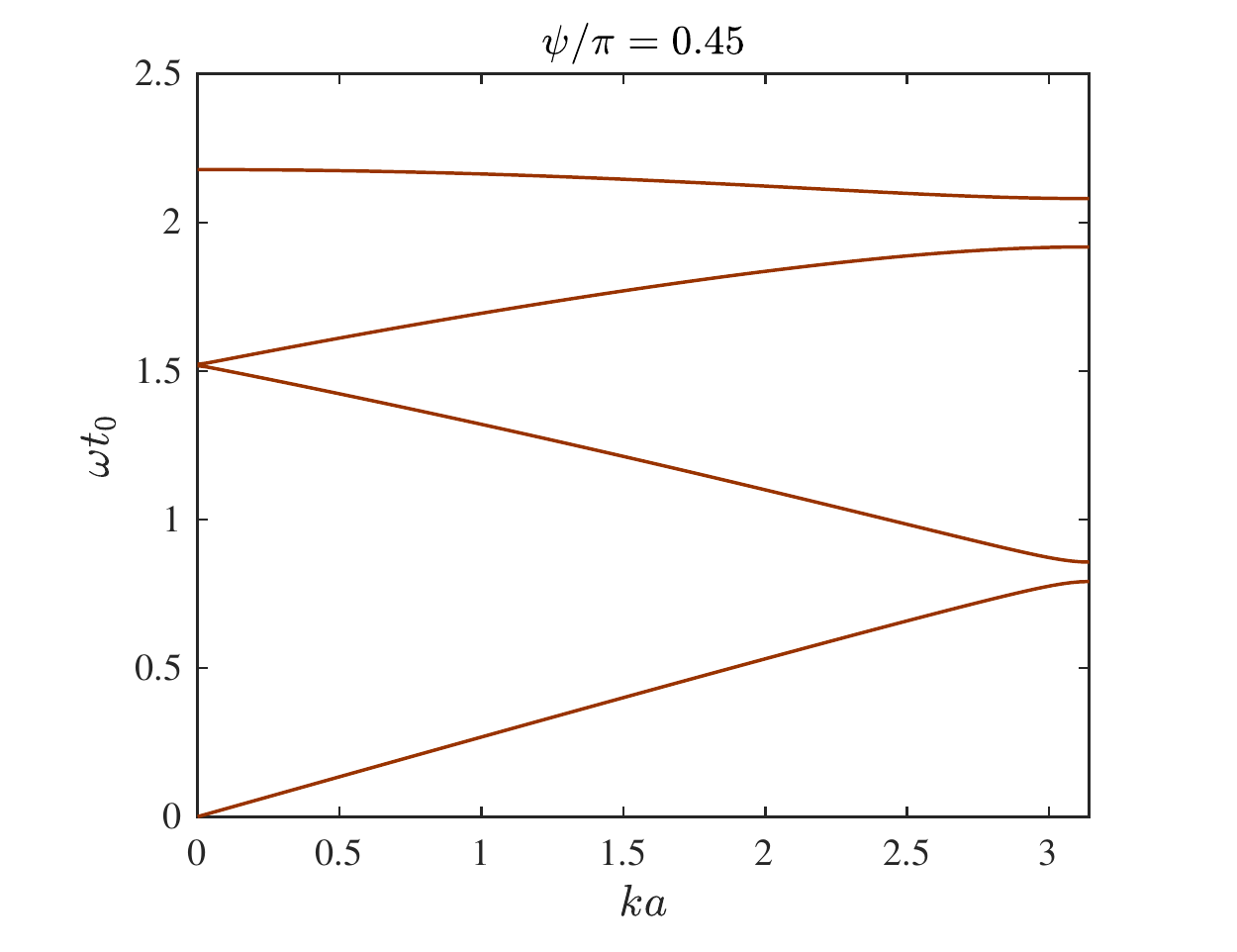}
\caption{Analytical dispersion relation of the compressed granular chain for an infinite number of particles, in the case $\psi/\pi = 0.45$.}
\label{fig:ChainDispersionRelation}
\end{figure}

In order to obtain analytical insight into the frequency response of our granular crystal with four-particle unit cells, we derive in Appendix \ref{app:TheoryCompressedChain} the dispersion relation of the system for an infinite number of particles, which relates the wavenumber $k$ of propagating sound waves to their frequency $\omega$. The dispersion relation is displayed in nondimensional form for the case $\psi/\pi = 0.45$ in Figure \ref{fig:ChainDispersionRelation}, where $k$ is normalized by the equilibrium length $a$ of each unit cell, and the corresponding $\omega$ is normalized by the $\psi$-independent time scale $t_0 = \sqrt{m/\kappa_3}$, with $m$ the mass of each sphere.
Compared with the three-particle unit cell studied in \cite{boechler2011}, we report the emergence of an additional fourth band of propagating frequencies above the usual acoustic and optical bands. As a consequence, our chain of four-particle unit cells inherits three bands of forbidden frequencies, or band gaps, in which sound waves decay exponentially and cannot propagate along the chain. The second band-gap, however, has negligible width for the material properties that we consider here.

\begin{figure}[tb!]
\centering
\includegraphics[width=0.5\textwidth]{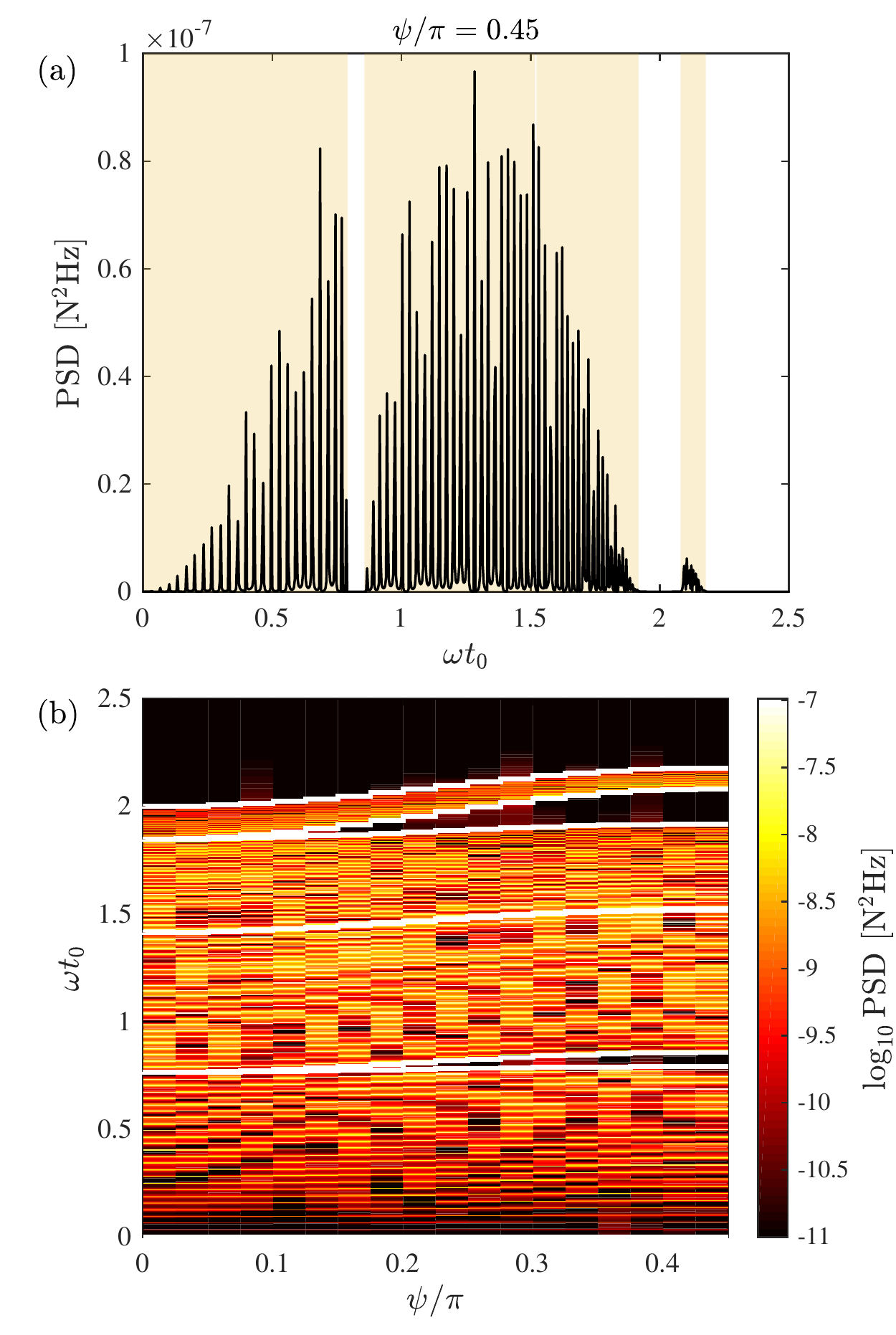}
\caption{Frequency response of the compressed granular chain. (a) Power spectral density of the force observed at the last grain for $\psi/\pi = 0.45$. The shaded regions correspond to the four bands of propagating frequencies predicted by the dispersion relation pictured in Figure \ref{fig:ChainDispersionRelation}. (b) Power spectral density of the force observed at the last grain, plotted in log scale for a range of values of $\psi/\pi$. The white lines are the cut-off frequencies predicted by the analytical dispersion relation for an infinitely long chain.}
\label{fig:ChainSpectrum}
\end{figure}

We now demonstrate the tunability of these vibrational band gaps by rotation of the middle particles in each unit cell. First, we verify the agreement between the analytical band frequencies and the behavior of a finite-length chain composed of $102$ particles, which we simulate in our anisotropic DEM framework. A small initial velocity is assigned to the first sphere in the chain, reproducing the effect of an impact excitation, and the force felt by the last sphere is measured as a function of time. Figure \ref{fig:ChainSpectrum}(a) shows the resulting power spectral density for the case $\psi/\pi = 0.45$, with the shaded regions corresponding to the four bands of propagating frequencies predicted by the dispersion relation pictured in Figure \ref{fig:ChainDispersionRelation}.
We observe excellent agreement between the theoretical predictions and numerical results, with vanishing energy of the force spectrum in the band gap regions. Next, we investigate the tunability of these band gaps by repeating the same numerical experiment for a range of orientations $\psi$ of the middle particles in each unit cell. The resulting spectrum is displayed in Figure \ref{fig:ChainSpectrum}(b) as a filled contour plot where each column corresponds to a particular value of $\psi$, and demonstrates the adjustability of the band gaps by simple rotation of some of the particles. The white lines are the cut-off frequencies predicted by the dispersion relation and agree very well with the numerics. We note that the band gaps disappear as $\psi$ goes to zero, corresponding to the limiting case of a uniform chain. Finally, it is worth keeping in mind that the dimensional band gap frequencies are a function of the precompression force through the power $-1/6$ dependence of the time scale $t_0$ on $F_0$, inherited from the dependence of $\kappa_3$ on $F_0$ (see equation \eqref{eq:LinearizedStiffness} in Appendix \ref{app:TheoryCompressedChain}). As a consequence, the precompression force provides an additional control parameter to tune the band gaps, besides the angle $\psi$.

\section{Conclusions and perspectives}
\label{sec:Conclusions}

In this paper, we have introduced a method to resolve the normal force arising between two elastically anisotropic contacting bodies of arbitrary geometry with smooth and frictionless surfaces, with the aim of obtaining a contact law that can be easily implemented into a DEM code. We first presented a numerical procedure for the exact solution of the full linear elasticity equations, resulting in an exact anisotropic contact force law. The computational cost of this exact contact law precluded its direct implementation into a DEM code, and its dependence on four parameters at a time prevented the use of a look-up table of precomputed values. 

By shortening the form of the full Green's function used in the exact solution, we then derived two simplifications to the exact contact law. \textit{Both simplifications take the same form as the Hertzian contact law for isotropic bodies}, save for the dependence of the contact (or plane strain) modulus associated with each body on the relative orientation of the contact normal direction and on the full set of elastic constants of the body. The precise form of the contact modulus differs between the two simplifications. In both cases, the parameter dependence of the computationally expensive part was reduced from four in the exact contact law down to two, a significant reduction that enabled the implementation of these simplified laws into a DEM code through the use of two-dimensional look-up tables of precomputed values of the contact modulus over all possible contact directions. Remarkably, the first of the two simplifications, which we called the truncated contact law, exhibited excellent accuracy compared to its exact counterpart, with the relative error on the predicted force remaining near or below 1\% for a wide range of materials and surface geometries.

Next, we presented the implementation of the truncated contact law into a DEM code, which we leveraged to showcase two application examples in which elastic anisotropy of the particles induced changes in the macroscopic behavior of the system. The first example we considered was that of a static square-based pyramid of contacting single-crystal zirconia spheres. By changing the orientation of the particles, we demonstrated that the pressure at the base of the pyramid is affected by the anisotropy of the contact forces. We then studied the transmission of sound waves in a compressed chain of adjacent single-crystal zirconia spheres, known as a `monoatomic granular crystal'. We leveraged the orientation-dependence of the contact stiffnesses between adjacent spheres to achieve frequency filtering characteristics that normally belong to the realm of diatomic granular crystals (assembled from two different constituent particles). More precisely, we revealed through theory and numerical computations the emergence of band gaps in which sound frequencies are unable to propagate down the chain. These band gaps are tunable by mere rotation of the particles, which offers an attractive prospect for adoption of such anisotropic granular crystals in scenarios that demand real-time control.

The present work opens the door to two distinct avenues of research. The first concerns the extension of our anisotropic contact law to frictional bodies, which can support tangential surface tractions in contrast to the frictionless bodies that we have treated. In general, the tangential force $F_{ij}^t$ is related to the normal force $F_{ij}^n$ through Coulomb's law, $F_{ij}^t \le \mu F_{ij}^n$, with $\mu$ a friction coefficient \cite{luding2008}. In order to determine the magnitude of $F_{ij}^t$ in the static friction case $F_{ij}^t < \mu F_{ij}^n$ as well as the onset of the dynamic friction case $F_{ij}^t = \mu F_{ij}^n$, the tangential contact law is typically regularized through a virtual tangential spring in a fashion that was pioneered by \cite{cundall1979}. While such an approach can be readily combined with our anisotropic contact law, we mention that several authors \cite{tsuji1992,vu1999,di2005} have developed more rigorous extensions of the tangential contact law for isotropic bodies, based on the early work of \cite{mindlin1953}. In these studies, the tangential spring becomes nonlinear and its stiffness is related to the elastic constants of the material, much in the same way that Hertzian contact theory provides a normal contact force law that is connected to the material parameters. Unlike the Hertzian normal force law, these tangential force relations depend on normal force history of the contact, which may have complex extensions in the anisotropic case. Although highly non-trivial, a generalization of the aforementioned studies to the elastically anisotropic case would be very valuable.

The second avenue of research enabled by the anisotropic contact law concerns the effect of elastic anisotropy on the behavior of granular systems, both at the microscopic and macroscopic levels. Granular materials sustain external loads through force chains, which are, in turn, responsible for the mechanical response of the sample \cite{zhang2017}. Considerable efforts have therefore been devoted to their characterization from both experimental \cite{majmudar2005} and theoretical \cite{snoeijer2004} perspectives. Recently, Hurley \textit{et al.} \cite{hurley2016} measured the distribution of contact forces in an assembly of elastically anisotropic quartz grains undergoing a compression cycle, and discovered a surprising inverse relationship between macroscopic load and heterogeneity of the contact forces, despite the clear formation of force chains. Reproducing their experiment in a DEM simulation using our anisotropic contact law could possibly shed light on the potential role of anisotropy in explaining their observation. Another potential area of application outside the realm of granular materials is the mechanical behavior of rock, which can be modeled in the DEM by a heterogeneous material comprised of cemented grains whose contact force law includes both grain-based and cement-based contributions  \cite{potyondy2004,jing2007,cho2007}. Although the elastic component of the grain-based portion of the normal contact law is usually considered isotropic, crystalline rocks such as granite possess a microstructure consisting of individual crystals, and would therefore benefit from the incorporation of our elastically anisotropic normal force law.

\begin{acknowledgements}
The authors are grateful to Shashank Agarwal for providing help with the finite-element simulations performed in ABAQUS. The authors acknowledge support from the US Army Institute for Soldier Nanotechnologies.
\end{acknowledgements}

\appendix

\section{Coordinate systems and transformations}
\label{app:CoordinateTransformations}

In this appendix, we introduce coordinate transformation matrices between the various reference frames that are utilized, which will come in handy when we describe the implementation of the contact force law in the following appendices. Recall Figure \ref{fig:GeometryBody}(a), which shows the two contacting bodies introduced in Figure \ref{fig:Geometry}, this time viewed from the global (laboratory) reference frame which is defined by the set of coordinates $(\mathcal{X}_1,\mathcal{X}_2,\mathcal{X}_3)$. In Section \ref{sec:AnisotropicBodies}, we have introduced a local set of coordinates $(X_1^B,X_2^B,X_3^B)$ that is oriented along the material structure of a given body $B$, rotating with it at all times. Finally, we also need to consider the set of coordinates $(x,y,z)$ aligned with the contact normal and tangent plane directions. Before proceeding further, we introduce three sets of orthonormal basis vectors: 
\begin{itemize}
\item $(\mathbf{e}_1^\mathcal{X}, \mathbf{e}_2^\mathcal{X}, \mathbf{e}_3^\mathcal{X})$, for the global coordinate system $(\mathcal{X}_1,\mathcal{X}_2,\mathcal{X}_3)$,
\item $(\mathbf{e}_1^X, \mathbf{e}_2^X, \mathbf{e}_3^X)$, for the body-centric coordinate system $(X_1^B,X_2^B,X_3^B)$,
\item $(\mathbf{e}_1^x, \mathbf{e}_2^x, \mathbf{e}_3^x)$, for the contact coordinate system $(x,y,z)$.
\end{itemize}

The orientation of body $B$ -- defined by coordinates $(X_1^B,X_2^B,X_3^B)$ -- with respect to the global reference frame $(\mathcal{X}_1,\mathcal{X}_2,\mathcal{X}_3)$ can be parameterized by a rotation matrix $[\mathbf{R}^B]$ whose components are defined by $R_{ij}^B = \mathbf{e}_i^\mathcal{X} \cdot \mathbf{e}_j^X$. (Appendix \ref{app:RelationshipEulerRotation} provides a relationship between these rotation matrices and the Euler parameters introduced in Section \ref{sec:EquationsMotion} to characterize the orientation of particles in the DEM code.) Further, we also introduce a coordinate transformation matrix $[\mathbf{Q}^B]$ from the contact basis to the body-centric basis, with elements given by $Q_{ij}^B = \mathbf{e}_i^X \cdot \mathbf{e}_j^x$.

With this in hand, one can relate the components of a vector $\mathbf{v}$ in the global or contact bases to its components in the local basis of body $B$ as 
\begin{subequations}
\begin{align}
\{\mathbf{v}\}^X &= [\mathbf{R}^B]^\mathsf{T} \{\mathbf{v}\}^\mathcal{X}, \label{eq:FromGlobalToBody} \\
\{\mathbf{v}\}^X &= [\mathbf{Q}^B] \{\mathbf{v}\}^x, \label{eq:FromContactToBody}
\end{align}
\end{subequations}
where the superscripts $X$, $\mathcal{X}$, and $x$ denote the components in the body-centric, global, and contact bases, respectively.

\section{Calculation of the Green's function}
\label{app:CalculationGreenFunction}

Algorithm \ref{alg:CalculationGreenFunction} presents a numerical procedure for computing the Green's function $h^B(\theta;0)$ introduced in Section \ref{sec:AnisotropicBodies}, which is a function of the elasticity tensor $\mathbb{C}^B$ as well as the relative orientation of the contact $(x,y,z)$ basis with respect to the body-centric $(X_1^B,X_2^B,X_3^B)$ basis. The latter is parameterized by the coordinate transformation matrix $[\mathbf{Q}^B]$ introduced in Appendix \ref{app:CoordinateTransformations}. Hereafter, we offer some complementary information on the algorithm. For the polar orientation $\theta$ and $\phi = 0$, the coordinates of the unit vectors $\mathbf{r}$ and $\mathbf{s}$ introduced in \eqref{eq:GreenUnitVectors} are given in the $(x,y,z)$ contact basis by
\begin{subequations}
\begin{align}
\{\mathbf{r}\}^x &= (\cos \gamma \sin \theta, - \cos \gamma \cos \theta, - \sin \gamma)^\mathsf{T}, \\
\{\mathbf{s}\}^x &= (-\sin \gamma \sin \theta, \sin \gamma \cos \theta, - \cos \gamma)^\mathsf{T},
\end{align}
\label{eq:GreenUnitVectors}%
\end{subequations}
which is used in line 3. In line 9, we have used the fact that the unit normal $\mathbf{n}$ is related to the basis vector $\mathbf{e}_3^x$ as $\mathbf{n} = -\mathbf{e}_3^x$; therefore its coordinates in the $(X_1^B,X_2^B,X_3^B)$ basis are given by $n_i^B = -Q_{i3}^B$. In practice, we discretize the integrals and iterate the \textbf{for} loops on lines 1 and 2 over $100$ values of $\theta$ and $\gamma$, equispaced between $0$ and $2\pi$.

\begin{algorithm}
\DontPrintSemicolon
\SetAlgoLined
\KwInput{Coordinate transformation matrix $[\mathbf{Q}^B]$ from $(x,y,z)$ basis to $(X_1^B,X_2^B,X_3^B)$ basis, components of elasticity tensor $\mathbb{C}^B$ in $(X_1^B,X_2^B,X_3^B)$ basis}
\For{$\theta = 0$ \KwTo $2\pi$}
{
	\For{$\gamma = 0$ \KwTo $2\pi$}
	{
		Calculate coordinates of $\mathbf{r}$, $\mathbf{s}$ in $(x,y,z)$ basis with \eqref{eq:GreenUnitVectors}\;
		Transform coordinates of $\mathbf{r}$, $\mathbf{s}$ to $(X_1^B,X_2^B,X_3^B)$ basis using \eqref{eq:FromContactToBody} with $[\mathbf{Q}^B]$\;
		Compute integrand of \eqref{eq:GreenIntegral} using \eqref{eq:GreenIntegralMatrices}\;
	}
	Perform integral in \eqref{eq:GreenIntegral} to get $G_{ij}(\theta;0)$\;
}
Using \eqref{eq:GreenFunctionAnisotropicPolar}, compute $h^B(\theta;0) \gets Q_{k3}^B G_{km}^{-1}(\theta;0) Q_{m3}^B$\;
\KwOutput{Green's function $h^B(\theta;0)$}
\caption{Calculation of the Green's function $h^B(\theta;0)$}
\label{alg:CalculationGreenFunction}
\end{algorithm}

\section{Solution strategy for $e$ and $\phi$}
\label{app:EccentricityPhaseSolution}

We describe our strategy to solve numerically the coupled equations \eqref{eq:OffsetAngleEccentricityRelations} for the eccentricity $e$ and phase angle $\phi$. First, we recast these equations as a minimization problem for the objective function $J(e,\phi) = \log (f_1^2(e,\phi) +f_2^2(e,\phi) )$, where $f_1(e,\phi)$ and $f_2(e,\phi)$ denote respectively the left-hand-sides of \eqref{eq:OffsetAngleEccentricityRelation1} and \eqref{eq:OffsetAngleEccentricityRelation2}. We then perform a global search for the minimum of $J$ on a coarse grid of values in the range $e \in [0,0.8]$ and $\phi \in [-\pi/2,\pi/2]$, and feed the resulting value as an initial condition to a gradient-based constrained optimization solver. We use MATLAB's \texttt{fmincon} function, which implements an interior-point algorithm, and constrain the search over the region $e \in [0,1]$. Since the objective function is $2\pi$-periodic in the $\phi$-direction, we have found that the optimization procedure is more robust when we leave $\phi$ unconstrained, and bring its value back to the interval $[-\pi/2,\pi/2]$ once the algorithm has converged.

\section{Calculation of a look-up table for $\tilde{E}_*^B$}
\label{app:CalculationPSM}

Given a material represented through its elasticity tensor $\mathbb{C}^B$, Algorithm \ref{alg:CalculationPSM} describes a numerical procedure for computing a look-up table of values of the equivalent plane strain modulus $\tilde{E}_*^B(\alpha^B,\beta^B)$ defined in Section \ref{sec:Approximation1}, for all possible orientations $\alpha^B \in [0,2\pi]$ and $\beta^B \in [0,\pi]$. We emphasize that the look-up table, $[\tilde{E}_*](\cdot,\cdot\,;\mathbb{C}^B)$, is purely a function of the elasticity tensor $\mathbb{C}^B$. As a result, a given simulation simply requires one look-up table per material present in the system. In the common case where all particles are made of the same material, only one such look-up table needs to be precomputed and stored.
In our implementation, we have used $100$ equispaced values for $\alpha^B$ and 50 for $\beta^B$. In line 7, the computation of the constant term of the Fourier series may be performed efficiently through the average $a_0^B = (2\pi)^{-1} \int_0^{2\pi} h^B(\theta;0) d\theta$. Note that the orientation of the unit vectors $\mathbf{u}$, $\mathbf{v}$ selected in line 4 of Algorithm \ref{alg:CalculationPSM} is inconsequential since only the mean component of the Green's function $h^B(\theta;0)$ is used. 

\begin{algorithm}
\DontPrintSemicolon
\SetAlgoLined
\KwInput{Components of elasticity tensor $\mathbb{C}^B$ in $(X_1^B,X_2^B,X_3^B)$ basis}
\For{$\alpha^B = 0$ \KwTo $2\pi$}
{
	\For{$\beta^B = 0$ \KwTo $\pi$}
	{
		Use \eqref{eq:FromEulerToComponents} to construct $\mathbf{n}$ from Euler angles $(\alpha^B,\beta^B)$\;
		Construct $\mathbf{u}$ and $\mathbf{v}$ such that $(\mathbf{u},\mathbf{v},\mathbf{n})$ forms an orthonormal basis\;
		Build the coordinate transformation matrix $[\mathbf{Q}^B] \gets [\{\mathbf{u}\}^X,\{\mathbf{v}\}^X,\{\mathbf{n}\}^X]$\;
		Call Algorithm \ref{alg:CalculationGreenFunction} using $[\mathbf{Q}^B]$ and $\mathbb{C}^B$ to get the Green's function $h^B(\theta;0)$\;
		Calculate $[\tilde{E}_*](\alpha^B,\beta^B;\mathbb{C}^B)$ from $h^B(\theta;0)$ with \eqref{eq:EquivalentPlaneStrainModulus}\;
	}
}
\KwOutput{Look-up table $[\tilde{E}_*](\cdot,\cdot\,;\mathbb{C}^B)$}
\caption{Calculation of a look-up table of precomputed values of $\tilde{E}_*^B$}
\label{alg:CalculationPSM}
\end{algorithm}

\section{Retrieving $\tilde{E}_*^c$ from the look-up table}
\label{app:CalculationCompositePSM}

We outline in Algorithm \ref{alg:CalculationCompositePSM} a procedure for retrieving the composite plain strain modulus $\tilde{E}_*^c$ between two contacting bodies $B_1$ and $B_2$ from their orientations and the look-up table(s) precomputed by Algorithm \ref{alg:CalculationPSM}. (Appendix \ref{app:RelationshipEulerRotation} presents formulae for obtaining the rotations matrices $[\mathbf{R}^{B_1}]$ and $[\mathbf{R}^{B_2}]$ characterizing the orientations of $B_1$ and $B_2$ from the Euler parameters utilized in Section \ref{sec:EquationsMotion}.) The algorithm is outlined for the general case where $B_1$ and $B_2$ are made of different materials with elasticity tensors $\mathbb{C}^{B_1}$ and $\mathbb{C}^{B_2}$, requiring the passage of two look-up tables as an input, one corresponding to each material. Note however that if $B_1$ and $B_2$ are made of the same material, then only one look-up table is required. As pointed out in Section \ref{sec:AnisotropicBodies}, the Green's function \eqref{eq:GreenFunctionAnisotropic}, and therefore the plane strain modulus $\tilde{E}_*^B$, are blind to the sign of the contact normal $\mathbf{n}$. Thus, we use in line 2 the same $\mathbf{n}$ to define the components of the contact normal direction in the reference frames of both bodies.

\begin{algorithm}
\DontPrintSemicolon
\SetAlgoLined
\KwInput{Rotation matrices $[\mathbf{R}^{B_1}]$ and $[\mathbf{R}^{B_2}]$ describing the orientations of bodies $B_1$ and $B_2$, components $\{\mathbf{n}\}^\mathcal{X}$ of contact normal direction $\mathbf{n}$ in global basis, look-up tables $[\tilde{E}_*](\cdot,\cdot\,;\mathbb{C}^{B_1})$ and $[\tilde{E}_*](\cdot,\cdot\,;\mathbb{C}^{B_2})$}
\For{$B = B_1,B_2$}
{
	Transform the coordinates of $\mathbf{n}$ from the global to the body's local $(X_1^B,X_2^B,X_3^B)$ basis using \eqref{eq:FromGlobalToBody} with $[\mathbf{R}^B]$\;
	Convert these coordinates to Euler angles $(\alpha^B,\beta^B)$ using \eqref{eq:FromComponentsToEuler}\;
	Use $(\alpha^B,\beta^B)$ to interpolate $\tilde{E}_*^B$ from the look-up table $[\tilde{E}_*](\cdot,\cdot\,;\mathbb{C}^B)$
}
Calculate $\tilde{E}_*^c$ using \eqref{eq:EquivalentCompositePlaneStrainModulus}\;
\KwOutput{Composite plain strain modulus $\tilde{E}_*^c$}
\caption{Retrieving the composite plain strain modulus $\tilde{E}_*^c$ }
\label{alg:CalculationCompositePSM}
\end{algorithm}

\section{Geometric features of the exact contact law}
\label{app:GeometricFeaturesExactHertzianSolution}

\begin{figure*}
\centering
\includegraphics[width=\textwidth]{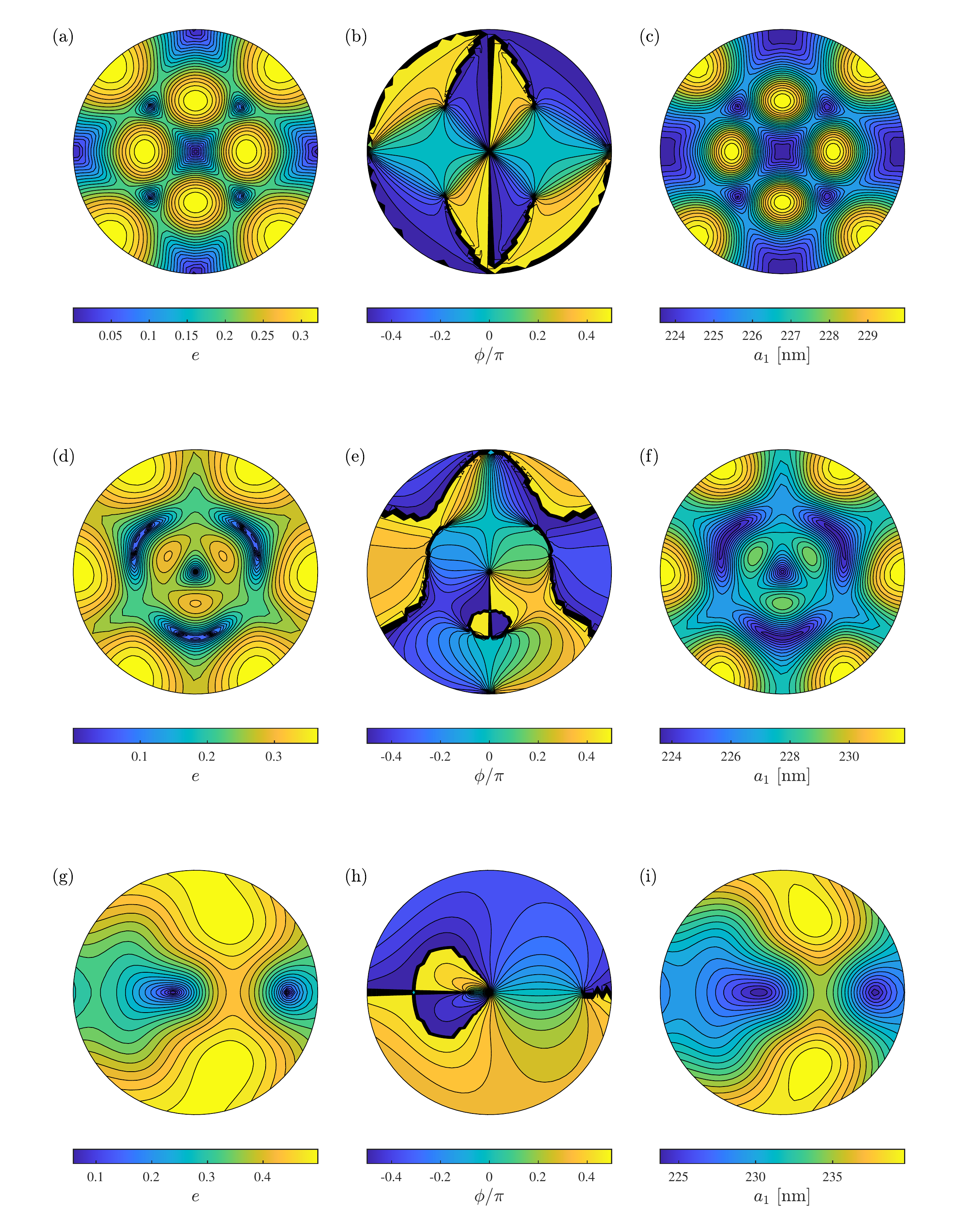}
\caption{Polar visualizations of the eccentricity $e$, orientation $\phi$, and semi-major axis length $a_1$ of the contact area predicted by the exact solution for iron (a,b,c), quartz (d,e,f), and zirconia (g,h,i), under indentation depth $\delta = 100 \, \text{nm}$ and gap function coefficients $A = B = 1 \, \mu\text{m}^{-1}$.}
\label{fig:ExactForceDetails_BoA1}
\end{figure*}

\begin{figure*}
\centering
\includegraphics[width=\textwidth]{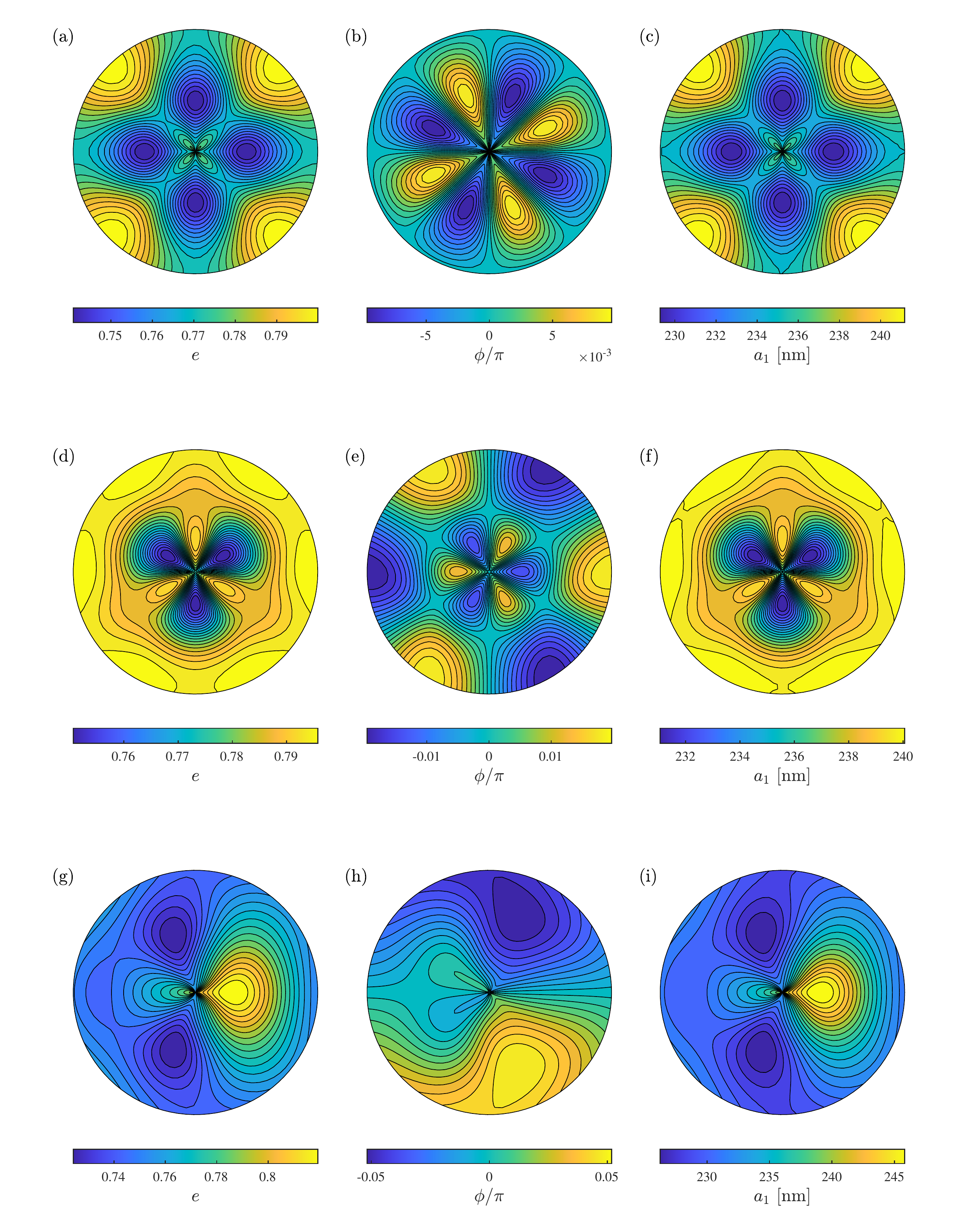}
\caption{Polar visualizations of the eccentricity $e$, orientation $\phi$, and semi-major axis length $a_1$ of the contact area predicted by the exact solution for iron (a,b,c), quartz (d,e,f), and zirconia (g,h,i), under indentation depth $\delta = 100 \, \text{nm}$ and gap function coefficients $A = 1 \, \mu\text{m}^{-1}$ and $B = 2 \, \mu\text{m}^{-1}$.}
\label{fig:ExactForceDetails_BoA2}
\end{figure*}

Here, we provide further details on the geometric features of the exact contact law for the materials and indentation parameters considered in Section \ref{sec:PolarVisualizations}. More specifically, we show polar visualizations of the eccentricity $e$, orientation $\phi$, and semi-major axis length $a_1$ of the contact area incurred by an indentation depth $\delta = 100 \, \text{nm}$, for a circular gap function ($A = B = 1 \, \mu\text{m}^{-1}$) in Figure \ref{fig:ExactForceDetails_BoA1} and an elliptic gap function ($A = 1 \, \mu\text{m}^{-1}$, $B = 2 \, \mu\text{m}^{-1}$) in Figure \ref{fig:ExactForceDetails_BoA2}. 
In both figures, (a,b,c) correspond to iron, (d,e,f) to quartz, and (g,h,i) to zirconia. Note that the discontinuities of the $\phi$ field appearing in Figure \ref{fig:ExactForceDetails_BoA1} are merely a visual artefact; indeed, the orientations $\phi = \pi/2$ and $\phi = -\pi/2$ are effectively identical as can be inferred from Figure \ref{fig:Geometry}(c). Interestingly, the contact normal direction-dependence of the fields $e$, $\phi$, and $a_1$ undergoes drastic change as the gap function changes from circular to elliptic, while that of the normal force $F$ remains relatively unaffected, as was shown in Figures \ref{fig:ForceComparison_BoA1} and \ref{fig:ForceComparison_BoA2}.

\section{Relationship between Euler parameters and rotation matrices}
\label{app:RelationshipEulerRotation}

In the DEM code presented in Section \ref{sec:Applications}, the orientation of a given particle $i$ is described using Euler parameters, or unit quaternions, $\{\bm{\epsilon}_i\}$ since their time integration from the angular velocity $\{\bm{\omega}_i\}$ is straightforward. In Appendices \ref{app:CoordinateTransformations} and \ref{app:CalculationCompositePSM}, however, the orientation is specified by a rotation matrix $[\mathbf{R}^{B_i}]$ (with $B_i$ referring to particle $i$), since the latter can be utilized to transform vector components from the global to the body-centric coordinate systems. Here, we specify the simple relationship that exists between the two representations. Dropping the index $i$, the rotation matrix is given by the Euler parameters as \cite{haug1989}
\begin{equation}
[\mathbf{R}^B] = 2 \left[
\begin{array}{ccc}
\epsilon_0^2 + \epsilon_1^2 -1/2 & \epsilon_1 \epsilon_2 - \epsilon_0 \epsilon_3 & \epsilon_1 \epsilon_3 + \epsilon_0 \epsilon_2 \\
\epsilon_1 \epsilon_2 + \epsilon_0 \epsilon_3 & \epsilon_0^2 + \epsilon_2^2 -1/2 & \epsilon_2 \epsilon_3 - \epsilon_0 \epsilon_1 \\
\epsilon_1 \epsilon_3 - \epsilon_0 \epsilon_2 & \epsilon_2 \epsilon_3 + \epsilon_0 \epsilon_1 & \epsilon_0^2 + \epsilon_3^2 - 1/2
\end{array}
\right],
\end{equation}
while the Euler parameters are given by the rotation matrix as
\begin{subequations}
\begin{align}
\epsilon_0^2 &= \frac{\text{tr} [\mathbf{R}^B]+1}{4}, \\
\epsilon_1 &= \frac{R_{32}^B-R_{23}^B}{4\epsilon_0}, \\
\epsilon_2 &= \frac{R_{13}^B-R_{31}^B}{4\epsilon_0}, \\
\epsilon_3 &= \frac{R_{21}^B-R_{12}^B}{4\epsilon_0}.
\end{align}
\end{subequations}
Note that the quadratic equation for $\epsilon_0$ possesses two roots, and the choice of a particular root also affects the signs of $\epsilon_1$, $\epsilon_2$, and $\epsilon_3$. Since the elements of $[\mathbf{R}^B]$ are quadratic in the Euler parameters, either root may be selected for $\epsilon_0$ and still define the same physical orientation of the body.

\section{Further details on the DEM implementation}
\label{app:DetailsDEM}

We provide additional details regarding our DEM code. We consider spherical zirconia particles with density $m = 5680 \, \mathrm{kg}/\mathrm{m}^3$ and uniform diameter $d = 1 \, \mathrm{cm}$. The elastic part of the normal force is calculated with our anisotropic contact law, using the elastic constants of zirconia given in Section \ref{sec:ComparisonContactForceLaws}. The viscous part is given a damping parameter $\gamma_n =  200 \, \mathrm{Ns}/\mathrm{m}$ in Section \ref{sec:StaticForceDistributionPyramid}, and $\gamma_n = 0$ in Section \ref{sec:SoundTransmissionGranularChain}. The code is implemented in MATLAB and utilizes a semi-implicit Euler method to evolve \eqref{eq:EvolutionEulerParameters} and \eqref{eq:EvolutionVelocities}. The linear and angular velocities are first integrated explicitly, following which the positions and orientations are integrated using the new (end-of-time-step) linear and angular velocities. We use a time step $\Delta t = 10^{-6} \, \mathrm{s}$.

\section{Theoretical analysis of the compressed chain}
\label{app:TheoryCompressedChain}

In this appendix, we derive the dispersion relation of the compressed chain of particles investigated in Section \ref{sec:SoundTransmissionGranularChain}. Our derivation follows the exposition of \cite{herbold2009} and \cite{boechler2011}, extending the latter to the present case of a four-particle unit cell. First, consider the force that is generated between any two particles $i$ and $j$ in the chain as a result of both the static force $F_0$ and the dynamic displacement of the particles. Following our contact force law \eqref{eq:IsotropicForceApproxCircular}, this force reads
\begin{equation}
F_{ij} = K_{ij} (\delta_{ij}^0 + \delta_{ij})^{3/2},
\label{eq:ForceOverlapRelationship}
\end{equation}
where $\delta_{ij}^0$, $\delta_{ij}$ are the overlaps between particles $i$ and $j$ due respectively to the static and dynamic force, and $K_{ij}$ is the nonlinear contact stiffness between particles $i$ and $j$, defined as
\begin{equation}
K_{ij} = \frac{2}{3} \tilde{E}_*^c(\alpha_i,\beta_i,\alpha_j,\beta_j) d^{1/2},
\end{equation}
with $\alpha_i$, $\beta_i$, $\alpha_j$, $\beta_j$ the Euler angles describing the orientations of particles $i$ and $j$ with respect to the contact normal direction (which is parallel to the chain axis), and $d$ the uniform diameter of the particles. Assuming that $\delta_{ij} \gg \delta_{ij}^0$, the force-overlap relationship \eqref{eq:ForceOverlapRelationship} can be linearized about $\delta_{ij}^0$, leading to
\begin{align}
F_{ij} \simeq K_{ij} (\delta_{ij}^0)^{3/2} + \frac{3}{2} K_{ij} (\delta_{ij}^0)^{1/2} \delta_{ij} = F_0 + \kappa_{ij} \delta_{ij},
\label{eq:LinearizedForceOverlapRelationship}
\end{align}
where we have substituted the static force $F_0 = K_{ij} (\delta_{ij}^0)^{3/2}$ and defined the linearized stiffness $\kappa_{ij}$ between particles $i$ and $j$ as
\begin{equation}
\kappa_{ij} = \frac{3}{2} K_{ij} (\delta_{ij}^0)^{1/2} = \frac{3}{2} K_{ij}^{2/3} F_0^{1/3}.
\label{eq:LinearizedStiffness}
\end{equation}
Going back to Figure \ref{fig:ChainGeometry}(a), we recall that our particles are oriented in a way that gives rise to three different possible stiffnesses between any two particles. Letting these stiffnesses $\kappa_1$, $\kappa_2$, and $\kappa_3$, we can write the linearized governing equations for the infinitely long chain as
\begin{subequations}
\begin{gather}
m \ddot{u}_{4n-3} =  \kappa_3 (u_{4n-4}-u_{4n-3}) - \kappa_2 (u_{4n-3}-u_{4n-2}), \\
m \ddot{u}_{4n-2} =  \kappa_2 (u_{4n-3}-u_{4n-2}) - \kappa_1 (u_{4n-2}-u_{4n-1}), \\
m \ddot{u}_{4n-1} =  \kappa_1 (u_{4n-2}-u_{4n-1}) - \kappa_2 (u_{4n-1}-u_{4n}), \\
m \ddot{u}_{4n} =  \kappa_2 (u_{4n-1}-u_{4n-0}) - \kappa_3 (u_{4n}-u_{4n+1}),
\end{gather}
\label{eq:ChainGoverningEquations}%
\end{subequations}
where $m$ denotes the uniform mass of the particles, $u_i$ is the dynamic displacement of particle $i$ with respect to its static equilibrium position in the compressed chain, and $n$ is the index of the unit cell. These are wave equations on a lattice with periodicity equal to the static unit cell length $a = 4d-\delta_1^0-2\delta_2^0-\delta_3^0$, where $\delta_c^0$ refers to the static overlap at a contact with stiffness $\kappa_c$. We therefore express the solution as a Bloch wave expansion,
\begin{equation}
\{u_{4n-3},u_{4n-2},u_{4n-1},u_{4n}\} = \{U,V,W,X\} e^{i(kan-\omega t)},
\label{eq:BlochWave}
\end{equation}
where the wavenumber $k$ belongs to the first Brillouin zone, $[-\pi/a,\pi/a]$. In order to find the frequency $\omega$ corresponding to each $k$, we substitute the expansion \eqref{eq:BlochWave} into \eqref{eq:ChainGoverningEquations} and solve for a nontrivial solution. This results in the dispersion relation
\begin{equation}
m^4 \omega^8 + c_6 m^3 \omega^6 + c_4 m^2 \omega^4 + c_2 m \omega^2 + c_0 = 0,
\label{eq:DispersionRelation}
\end{equation}
where $c_6$, $c_4$, $c_2$, $c_0$ are functions of $\kappa_1$, $\kappa_2$, $\kappa_3$ as follows:
\begin{subequations}
\begin{align}
c_6 &= -2(\kappa_1+2\kappa_2+\kappa_3), \\
c_4 &= -\kappa_1^2 - 2\kappa_2^2 - \kappa_3^2 + (\kappa_1+\kappa_2)^2 + (\kappa_2+\kappa_3)^2 \nonumber \\
&\quad + 4(\kappa_1+\kappa_2)(\kappa_2+\kappa_3), \\
c_2 &= 2(\kappa_1+\kappa_2)(\kappa_2^2+\kappa_3^2) + 2(\kappa_2+\kappa_3)(\kappa_1^2+\kappa_2^2) \nonumber \\
&\quad - 2(\kappa_1+\kappa_2)(\kappa_2+\kappa_3)^2 - 2(\kappa_2+\kappa_3)(\kappa_1+\kappa_2)^2, \\
c_0 &= \kappa_1^2\kappa_3^2 + \kappa_2^4 + (\kappa_1+\kappa_2)^2(\kappa_2+\kappa_3)^2 -\kappa_1^2(\kappa_2+\kappa_3)^2 \nonumber \\
&\quad - \kappa_3^2(\kappa_1+\kappa_2)^2 - 2\kappa_1\kappa_2^2\kappa_3\cos ka = 0.
\end{align}
\end{subequations}
The dispersion relation \eqref{eq:DispersionRelation} possesses four $\omega$ solutions for every value of $k$, which are plotted in Figure \ref{fig:ChainDispersionRelation} in the range $k \in [0,\pi/a]$ due to the symmetry of $c_0$ with respect to $k=0$.

 \section*{Code availability}
A Python code implementing the truncated simplification of the contact force based on the approach described in Section \ref{sec:EfficientImplementationInDEM}, including the precomputation of a look-up table of plane strain modulus values, has been shared in an online repository at \url{https://github.com/smowlavi/AnisotropicGrains.git}.

%
 \section*{Conflict of interest}

 The authors declare that they have no conflict of interest.

\bibliographystyle{spmpsci}      

\end{document}